\documentclass[twocolumn]{emulateapj}
\pdfoutput=1
\usepackage{amssymb}
\usepackage{amsmath}
\usepackage{graphicx}
\usepackage{natbib}
\usepackage{bigints}
\usepackage{relsize}
\usepackage{color}

\newcommand{\targ}{2MJ0830+15~}
\newcommand{\redcol}{\color{black}}

\shorttitle{Discovery of the First Triple T-dwarf}
\shortauthors{Radigan et al.}

\begin{document}


\title{Discovery of a Visual T-dwarf Triple System and Binarity at the L/T Transition \footnote{T\lowercase{he data presented herein were obtained at the} W.M. K\lowercase{eck} O\lowercase{bservatory, which is operated as a scientific partnership among the} C\lowercase{alifornia} I\lowercase{nstitute of} T\lowercase{echnology, the} U\lowercase{niversity of} C\lowercase{alifornia and the} N\lowercase{ational} A\lowercase{eronautics and} S\lowercase{pace} A\lowercase{dministration}. T\lowercase{he} O\lowercase{bservatory was made possible by the generous financial support of the} W.M. K\lowercase{eck} F\lowercase{oundation.}}}


\author{Jacqueline Radigan\altaffilmark{1,2}, Ray Jayawardhana\altaffilmark{2}, David Lafreni{\`e}re\altaffilmark{3}, Trent J.\ Dupuy\altaffilmark{4,5}, Michael C. Liu\altaffilmark{6}, Alexander Scholz\altaffilmark{7}}
\email{radigan@stsci.edu}

\altaffiltext{1}{Space Telescope Science Institute, 3700 San Martin Drive, Baltimore, MD 21218}
\altaffiltext{2}{Department of Astronomy \& Astrophysics, University of Toronto, 50 St. George Street, Toronto, ON M5S~3H4, Canada}
\altaffiltext{3}{D{\'e}partement de Physique, Universit{\'e} de Montr{\'e}al, C.P. 6128 Succ. Centre-Ville, Montr{\'e}al,  QC H3C~3J7, Canada}
\altaffiltext{4}{Harvard-Smithsonian Center for Astrophysics, 60 Garden Street, Cambridge, MA 02138}
\altaffiltext{5}{Hubble Fellow}
\altaffiltext{6}{Institute for Astronomy, University of Hawaii, 2680 Woodlawn Drive, Honolulu, HI 96822}
\altaffiltext{7}{School of Cosmic Physics, Dublin Institute for Advanced Studies, 31 Fitzwilliam Place, Dublin 2, Ireland}


\begin{abstract}
We present new high contrast imaging of 8 L/T transition brown dwarfs using the NIRC2 camera on the Keck II telescope.  One of our targets, the T3.5 dwarf 2MASS J08381155$+$1511155, was resolved into a hierarchal triple with projected separations of 2.5$\pm$0.5\,AU and 27$\pm$5\,AU for the BC and A(BC) components respectively.  Resolved OSIRIS spectroscopy of the A(BC) components confirm that all system members are T dwarfs.  The system therefore constitutes the first triple T-dwarf system ever reported.  Using resolved photometry to model the integrated-light spectrum, we infer spectral types of T3$\pm$1, T3$\pm$1, and T4.5$\pm$1 for the A, B, and C components respectively. The uniformly brighter primary has a bluer $J-K_s$ color than the next faintest component, which may reflect a sensitive dependence of the L/T transition temperature on gravity,  or alternatively divergent cloud properties amongst components.  Relying on empirical trends and evolutionary models we infer a total system mass of 0.034-0.104\,$M_{\odot}$ for the BC components at ages of 0.3-3\,Gyr, which would imply a period of 12-21\,yr assuming the system semi-major axis to be similar to its projection.  We also infer differences in effective temperatures and surface gravities between components of no more than $\sim$150\,K and $\sim$0.1\,dex.  Given the similar physical properties of the components, the 2M0838$+$15 system provides a controlled sample for constraining the relative roles of effective temperature, surface gravity, and dust clouds in the poorly understood L/T transition regime.
For an age of 3\,Gyr we estimate a binding energy of $\sim$20$\times 10^{41}$\, erg for the wide A(BC) pair, which falls above the empirical minimum found for typical brown dwarf binaries, and suggests that the system may have been able to survive a dynamical ejection during formation.  Combining our imaging survey results with previous work we find an observed binary fraction of 4$/$18 or $22_{-8}^{+10}$\% for unresolved spectral types of L9-T4 at separations $\gtrsim 0.1$\arcsec.  This translates into a volume-corrected frequency of $13^{+7}_{-6}$\%, which is similar to values of $\sim$9-12\% reported outside the transition.   Our reported L/T transition binary fraction is roughly twice as large as the binary fraction of an equivalent L9-T4 sample selected from primary rather than unresolved spectral types ($6^{+6}_{-4}$\%), however this increase is not yet statistically significant and a larger sample is required to settle the issue.
\end{abstract}


\section{Introduction}
The L/T transition, roughly spanning L8-T5 spectral types, is characterized by dramatic spectral evolution at near-infrared wavelengths at a near-constant effective temperature of $\sim$1200\,K \citep{golimowski04,stephens09}.  Rather than following a sequence in effective temperature, evolution across the L/T transition follows the disappearance of condensate clouds from brown dwarf photospheres as progressively larger dust grains gravitationally settle faster than they can be replenished
\citep[e.g.][]{ackerman01,tsuji02,marley02,allard03,woitke03}.  As condensate opacity declines, the average $\tau=2/3$ surface moves to deeper, warmer atmospheric layers.    This effect is seen most clearly in at wavelengths $\sim$1\,$\mu$m ($Y$ and $J$-bands) for which clouds are a dominant opacity source.   From L8-T5 spectral types the $\sim$1\,$\mu$m flux increases by a factor as high as $\sim$2.5, \citep[``J-band brightening''; ][]{dahn02,tinney03,vrba04,dupuy12,faherty12}, as the dust opacity declines.
 The observed constancy in effective temperature across the L/T transition results from a coincidence of opposing phenomena:  cooling of the atmosphere is roughly offset by its increasing transparency to deeper layers \citep{saumon08}.
 The processes governing the dissipation and settling of condensates across the L/T transition remain poorly understood, with models generally predicting a more gradual change in cloud properties over a wider range of effective temperatures than is observed \citep[e.g.][]{tsuji03,marley02,allard03}.  It is debated as to what extent the rapid decrease in cloud opacity is due to global changes in cloud thickness and position, changing grain properties, increasing rates of grain sedimentation, or a decreasing cloud filling fraction over the brown dwarf's surface \citep[e.g.][]{burrows06,saumon08,burgasser02_lt}.  Recent observations of the highly variable L/T transition dwarfs SIMP0136 \citep{artigau09} and 2M2139+02 \citep{radigan12} suggest that heterogeneous cloud cover may indeed contribute to the declining condensate opacity in this regime. 
 
 Multiple systems have already played an important role in testing formation, atmosphere and evolutionary models for very low mass (VLM, $M_1\lesssim 0.1\,M_{\odot}$) and substellar objects in general \citep[e.g.][]{liu06,stassun07,burgasser07_bin,dupuy12}.  When attempting to understand the L/T transition, multiple systems are important in two major ways.   First, it has been suggested that there could be an unusually high binary fraction in this regime ($\sim$L9-T4 spectral types).  According to a population synthesis of \citet{burgasser07_ltbin}, this hypothesized increase reflects a dip in the luminosity function of bonafide single brown dwarfs at these spectral types (e.g. due to rapid evolution through this regime), in contrast to a roughly constant luminosity function for unresolved binaries whose integrated light spectral types mimic those of true transition objects.  Thus multiplicity searches at the L/T transition are important in order to determine the degree of binary contamination, and to identify contaminants.  Second, if L/T transition binaries are resolved into constituent parts in or straddling the L/T transition, they can provide strong tests of models where two free parameters, the age and metallicity of the system, are fixed.   For instance, the identification of ``flux-reversal'' binaries wherein the secondary component is brighter than the primary in the $J$ band has provided direct evidence that this brightening is a real evolutionary feature associated with the disappearance of dust clouds \citep[e.g.][]{liu06,looper08}.  Furthermore, in cases where masses can be constrained from visual orbits these systems can act as gravity benchmarks which will further our understanding of the dependence of cloud properties on surface gravity.
 
Here we present high contrast imaging observations of 8 brown dwarfs occupying the sparsely populated L/T transition, designed to search for hitherto undetected multiples.   One object, the T3.5 dwarf 2MASS J08381155$+$1511155 (2M0838$+$15 hereafter), was resolved into a triple system.  This discovery  constitutes the first triple T-dwarf ever reported.  In section \ref{sect:obs} we describe our target sample including the discovery of 2M0838$+$15, our observations and data reduction.  In section \ref{sect:triple} we analyze the resolved NIRC2 images and OSIRIS spectroscopy of the newly resolved 2M0838$+$15 system.  In section \ref{sect:survey} we describe the search for companions and detection limits around our entire sample.  In section \ref{sect:stats} we combine our results with those from previous surveys to infer a binary fraction for L9-T4 spectral types, and to test whether there is a statistically significant increase in binary frequencies inside the L/T transition.  In section \ref{sect:discussion} we summarize our major findings and discuss our results in the context of the L/T transition and substellar formation models.

\section{Observations and Data Reduction}
\label{sect:obs}
Here we present high contrast imaging observations of 8 L/T transition brown dwarfs (BDs) using the NIRC2 camera on the Keck telescope obtained on the night of 2010 January 8.  One target in our sample was resolved into a triple system, for which we obtained follow up imaging with NIRC2 on 22 March 2010 and partially resolved spectroscopy using OSIRIS on 2011 December 03.  These three data sets are described in the following subsections.

\subsection{Target Selection}\label{sect:targ}
Details for our NIRC2 imaging targets are provided in table \ref{tab:targets}.  The targets were selected to overlap with an L/T transition sample of brown dwarfs targeted for variability monitoring \citep{radigan11} where possible, while gaps in our program were filled by additional targets with late-L and T spectral types that were observable at low airmass at the time of our observations.  Objects were further culled based on not having previously been targeted by a high resolution imaging survey, and possessing a suitable tip-tilt star within 50\arcsec. One object fulfilling these criteria was an unpublished T dwarf, 2M0838$+$15, whose discovery and spectral confirmation is described in the following subsection. 

\subsubsection{Discovery and Spectral Confirmation of 2M0838$+$15}
The early T dwarf 2M0838$+$15 was discovered in 2008 from our own proper motion cross-match of 2MASS and SDSS catalogs \citep[described in][]{radigan08}, but remained unreported.  This source was independently discovered as a T-dwarf candidate in a crossmatch of 2MASS and WISE by \citet{aberasturi11}.

We obtained spectral confirmation for 2M0838$+$15 on 2008 March 01 using the SpeX Medium-Resolution Spectrograph \citep{rayner03} at NASA's Infrared Telescope Facility (IRTF).  Observations were made in the short slit (15\arcsec) prism mode (0.8-2.5\,\micron), with a 0.5\arcsec wide slit.  The seeing was 0.8\arcsec-0.9\arcsec.  We obtained ten 180\,s exposures consisting of 5 AB pairs with a nod step of 7\arcsec along the slit. For telluric and instrumental transmission correction the A0V star HD 79108 was observed immediately after the target at a similar air mass. Flat-fielding, background subtraction, spectrum extraction, wavelength calibration, and telluric correction were done using Spextool \citep{cushing04,vacca03}. 

\begin{figure*}
\epsscale{0.7}
\plotone{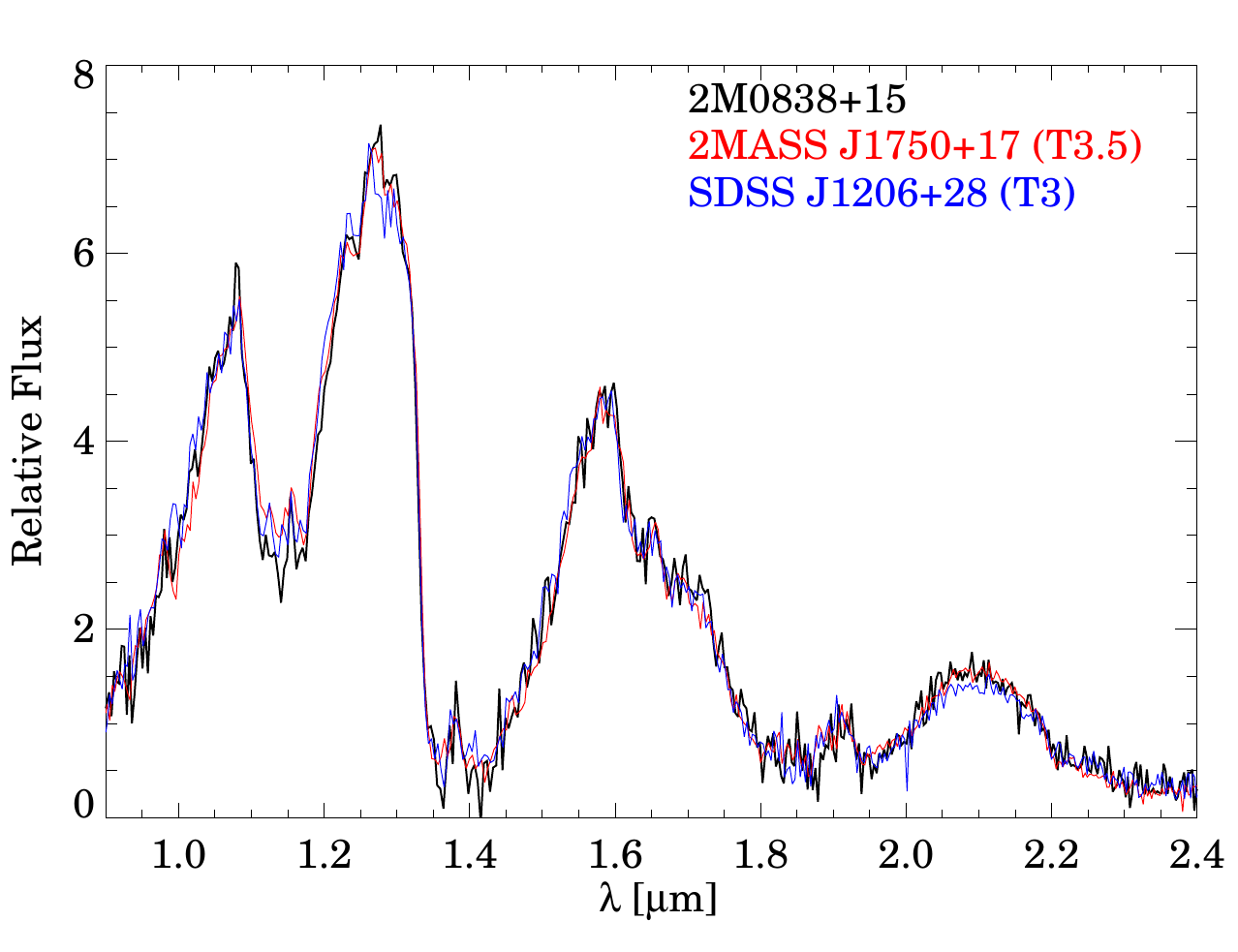}
\caption{Unresolved JHK spectrum of the 2M0838$+$15ABC system (black line), obtained using SpeX at the IRTF.  The best-fitting template from the SpeX Prism Library, the T3.5 dwarf SDSSp~J175032.96$+$175903.9 \citep{geballe02,burgasser06}, is over plotted as a spectral type reference (red).  The next best fitting template, the T3 dwarf  SDSS~J120602.51$+$281328.7 \citep{chiu06} is also shown. The camera position angle on the sky is 45.7$^{\circ}$.  \label{fig:spt_match}}
\end{figure*}

The unresolved spectrum for \targ is presented in figure \ref{fig:spt_match}.  Based on least squares fitting of spectral templates in the SpeX Prism Library to our data we derive a spectral type of T3.5$\pm$0.5 for the unresolved source.

\begin{figure*}
\epsscale{0.85}
\plotone{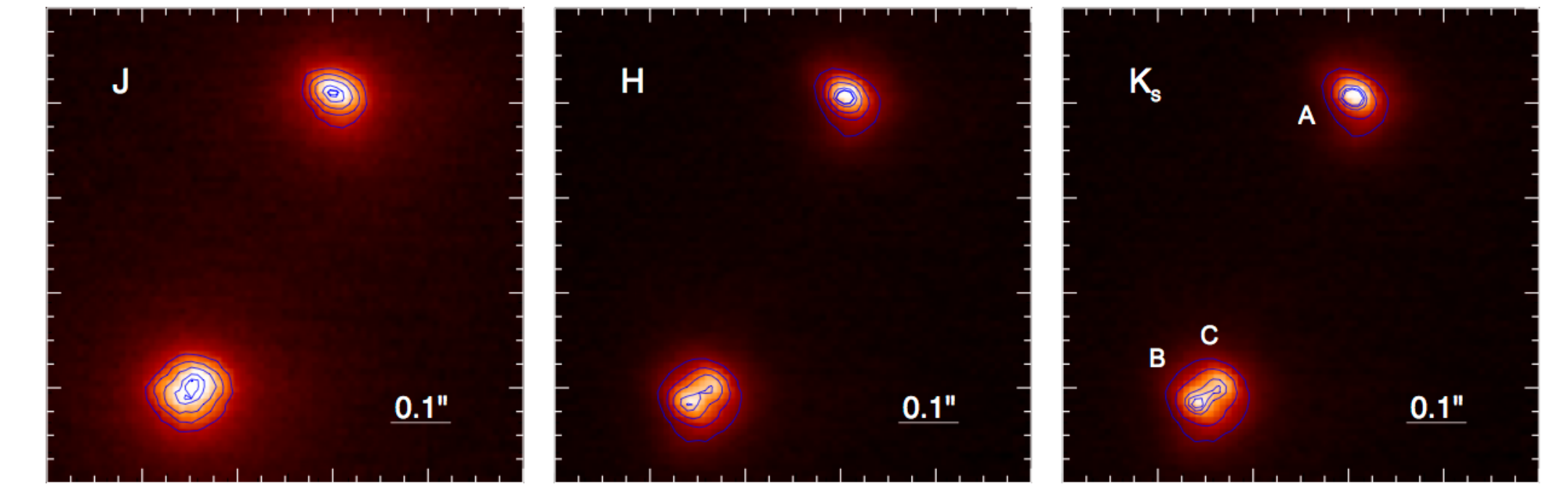}{
\caption{Reduced NIRC2 images in the $J$, $H$, and $K_s$ bands  (from left to right) of the 2M0838$+$15ABC system. \label{fig:jhk_mos}}}
\end{figure*}

Due to its status as an L/T transition object without previously reported high contrast imaging observations, 2M0838$+$15 was included in our LGS AO mini-survey of L/T transition BDs, described above in section \ref{sect:survey}.  Images obtained using the NIRC2 camera during the observations of 2010 January 08 revealed this source to be a hierarchal triple (figure \ref{fig:jhk_mos}) consisting of a widely separated A(BC) pair ($\sim$0.5\arcsec), with the BC component being further resolved into a tight double ($\sim$0.05\arcsec).

\subsection{NIRC2 Observations of L/T transition dwarfs}
\label{sect:nirc_obs}
Observations of 8 L/T transition BDs (see table \ref{tab:targets}) were obtained in the first half of the night of 2010 Jan 8 using the NIRC2 narrow camera on the Keck II telescope.  Due to the lack of bright natural guide stars within 20\arcsec of our targets, the laser guide star and an off axis tip-tilt (TT) star were used for adaptive optics (AO) corrections.  For each target we obtained 3 or more dithered images in the $K_s$ band.   We used a 3-point dither pattern, offsetting targets $\pm$2.5\arcsec~ from the center of the array in both x$-$ and y$-$ directions, excluding the bottom-left quadrant of the array which is significantly noisier than the others.  The median FWHM and Strehl ratios achieved in the $K_s$ band were 0.073\arcsec and  0.19 respectively.  Some targets were also observed in the $J$ and $H$ bands but with lower image quality.  Here, we only present multi-band data for the lone target in our sample that was resolved into a multiple system.  Details pertaining to observations of individual targets including airmass, exposure times, number of exposures, average FWHM and Strehl ratios, and TT star magnitudes and separations are provided in table \ref{tab:obs}.  

For calibration purposes ten to fifteen dome flat fields (with the lamp on and off) for each bandpass were taken before sunset.  Dark frames were obtained the afternoon after the observations for all but the longest exposure times (180\,s and 210\,s).  For exposure times without corresponding dark frames the longest exposure (120\,s) dark frames were simply scaled by integration time.  The dark frames for each exposure time were median combined to form master dark frames for each exposure time.  A master dark of the appropriate exposure time was then subtracted from all other science images.  The flat field frames for a given filter were median combined and the resultant lamp-off frames subtracted from the lamp-on frames to obtain a single high-signal-to-noise flat field for each filter.  All sciences images were divided by the flat field to correct for pixel-to-pixel variations in quantum efficiency.  For each science image of a given target in a given bandpass a sky frame was obtained by averaging together all other exposures wherein the target did not fall in the same quadrant of the array.   The iterative (3-$\sigma$ clipped) median of each sky frame was scaled to match the iterative median of the corresponding science frame in the target quadrant. The scaled sky frames were then subtracted from the science frames.  A bad pixel mask was constructed by identifying hot pixels in the dark frames and dead pixels in the flat field frames.  Additional bad pixels missed by this method were manually flagged.  Bad pixels located more than 2 FWHM away from the target were corrected by replacing their value with that of a 5$\times$5 pixel median filtered image.  Bad pixels falling within 2 FWHM of the target were interpolated using a second order surface interpolation of neighboring pixels.  All pixel interpolations in the vicinity of our targets were also examined by eye.  In addition to retaining the reduced individual exposures, all science images for a given filter and  target were  positionally cross-correlated against one another to determine relative sub-pixel offsets, interpolated onto a common grid, and stacked.  Individual exposures with the narrowest FWHMs were used to search for close companions as described in section \ref{sect:survey}, while the stacked images were used to place limits on the presence of well-separated faint companions.

\begin{figure*}
\epsscale{1.1}
\plotone{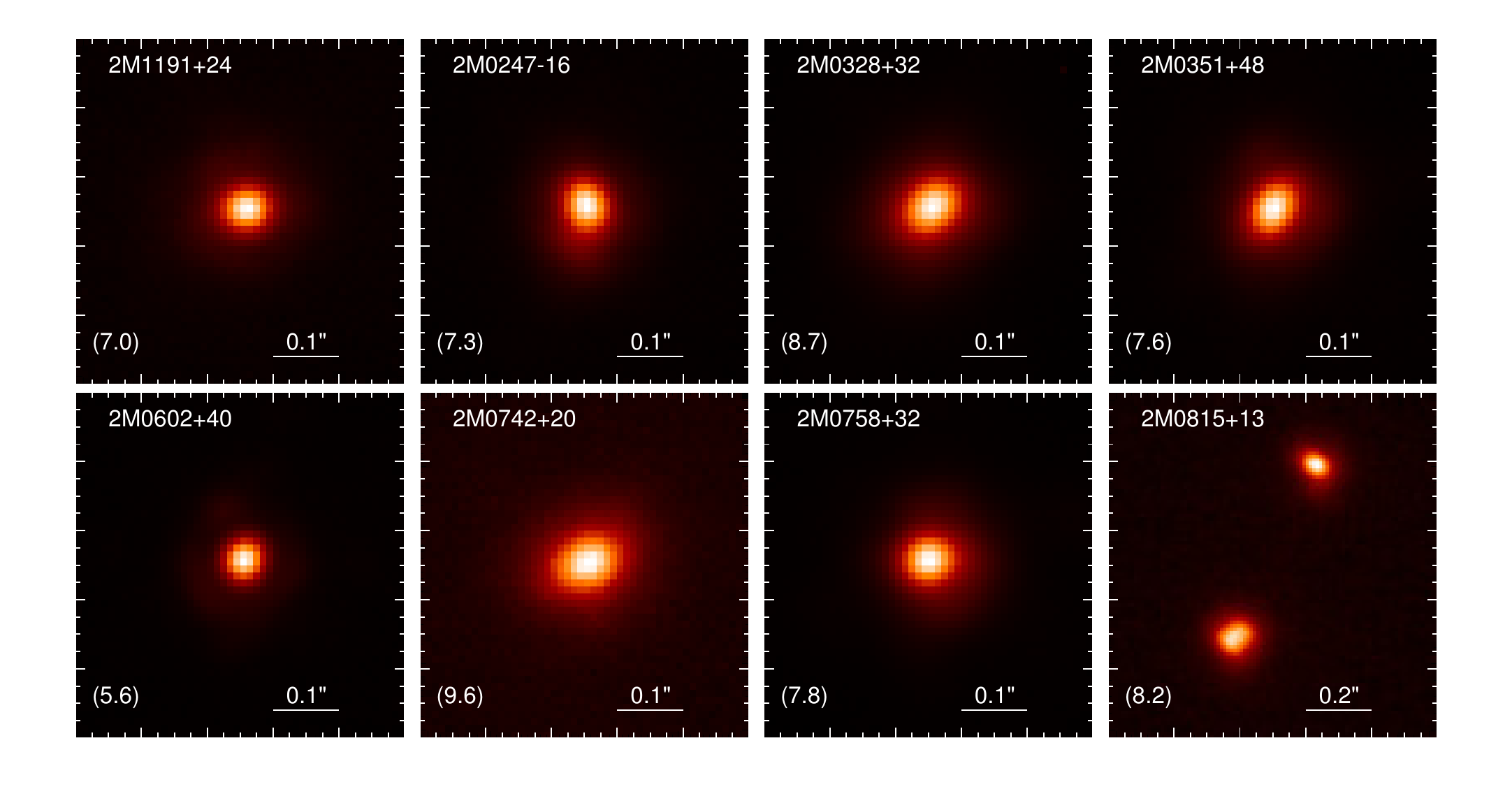}
\caption{Postage stamps ($K_s$ band) of all targets observed with NIRC2.  In several cases the PSF is elongated in the direction of the TT star.  Only a single target 2M0838$+$15 was resolved into a multiple system (bottom right).  The camera position angle on the sky is 45.7$^{\circ}$. The PSF FWHM is indicated in brackets.  \label{fig:thumbnails}}
\end{figure*}

Reduced images of our targets are shown in figure \ref{fig:thumbnails}.  One of eight targets, the T3.5 dwarf 2M0838$+$15, was resolved into a multiple system.  Images obtained revealed this source to be a hierarchal triple consisting of a widely separated A(BC) pair ($\sim$0.5\arcsec), with the BC component being further resolved into a tight double ($\sim$0.05\arcsec).  Examples of reduced and sky-subtracted  $J$, $H$, and $K_s$ images of the 2M0838$+$15 ABC system are shown in figure \ref{fig:jhk_mos}.  We analyze these images to obtain fluxes and system parameters in section \ref{sect:mpfit}

\subsection{OSIRIS spectroscopy of the 2M0838$+$15 system}
Spatially resolved spectra of the the 2M0838$+$15 components were obtained on 2011 Dec 03 in the latter half of the night using the OSIRIS integral field unit \citep{osiris} on the Keck II telescope.  Conditions were partially cloudy.  The extinction only dropped below 1 mag, permitting operation of the laser, in the latter quarter of the night.  The star USNO-A2.0 050-05806001, located 51.1\arcsec~ away, was used for TT corrections, and fell in a non-vignetted region of the guide camera field of view when the 2M0838$+$15 A(BC) system was aligned along the long axis of the OSIRIS spectrograph (a position angle of 19 degrees on the sky).  We had originally planned to use the finest 0.01\arcsec\,pix$^{-1}$ plate scale and obtain blank sky frames for sky subtraction.  However, because  of time lost due to clouds in the first half of our run, we instead opted to use the 0.035\arcsec\,pix$^{-1}$ scale which provided a slightly larger field of view and increased efficiency by allowing us to dither on-chip rather than requiring separate blank sky frames.  We obtained two AB pairs of spectra in the $H$-band with 15 minute and 5 minute exposures respectively, and one AB pair of $K$-band spectra with 6 min exposures.  

We observed the A0V star HD64586 and a probable K giant star BD+211974 for telluric correction before ($H$ band only) and after ($H$ and $K$ bands) the science observations, at similar airmass.  Due to bad weather and initial problems acquiring the science target, observations of the first telluric occurred over 2 hours ahead of the science exposures.  Unfortunately although our second telluric BD+211974 is listed in SIMBAD as an A0V star, we discovered upon obtaining a spectrum that it is more likely an M or K giant with strong CO absorption features in the $K$ band (although relatively featureless in $H$).   In order to use  BD+211974 for telluric correction, knowledge of its intrinsic spectrum is required.  To this end we obtained a spectrum of BD+211974 using SpeX at the IRTF in short wavelength cross dispersed mode on 9 Jun 2012, using the A0V star HD79108 for telluric correction.  The spectrum was reduced using SpeXtool \citep{cushing04,vacca03}.

The raw spectra were sky subtracted in AB pairs, wavelength calibrated and converted into 3D data cubes (2 spatial directions plus wavelength) using the OSIRIS data reduction pipeline (DRP) with the latest rectification matrices available (2010 July).   The pipeline also corrects for bias variations between detector output channels, crosstalk, electronic glitches, and attempts cosmic ray removal.  1-D spectra were extracted from the reduced data cubes via aperture photometry on the individual wavelength slices.  A circular aperture of 2.5 pixel radius centered on each of the A and BC components was used.  Residual sky levels were measured in annuli of 6 and 9 pixel inner and outer radii.
Extraction of the brighter standard star spectra were conducted using a larger 4-pixel radius aperture, and residual sky levels were found to be negligible.  For the A0V star HD64586 we used the XtellCorr software package described in \citet{vacca03} to obtain our telluric spectrum (including scaled hydrogen line removal, and division of the spectrum by a Vega template).  For the giant star BD+211974 we determined a telluric correction by dividing our non-corrected OSIRIS spectrum of BD+211974 by the fully corrected (i.e. intrinsic) SpeX spectrum.  In the $K$ band BD+211974 is our only option for telluric correction, while in the $H$ band 
both HD64586 and BD+211974 were available.  When applied to the $H$ band science data, we found that the latter provided a slightly cleaner correction and was thus adopted in our final reduction.   We verified the DRP wavelength solution by comparing a sky spectrum to a database of OH lines and found it to be good to within $\sim$10 \AA, which is more than sufficient for our purposes.  

\begin{figure*}
\epsscale{0.7}
\plotone{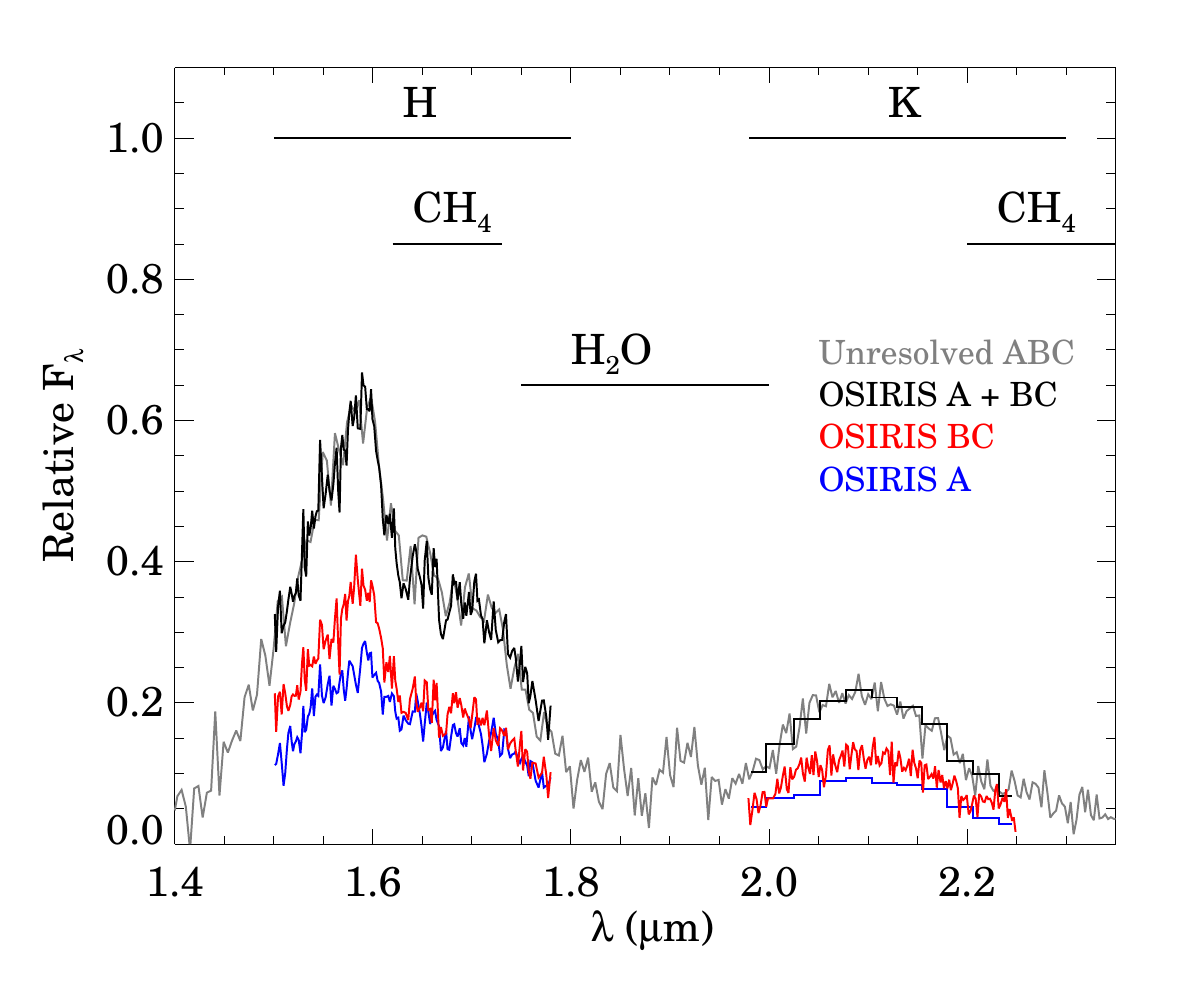}
\caption{Resolved $H$ and $K_s$ spectra of the A (blue line) and BC (red line) components obtained using the OSIRIS spectrograph on the Keck II telescope.  The sum of the A and BC contributions is plotted as a black line. The unresolved SpeX spectrum is shown for reference in grey and was used to scale the relative $H$ and $K_s$ contributions.  \label{fig:triple}}
\end{figure*}

The resultant $H$ and $K$ spectra for the A and BC components, in units of relative $F_{\lambda}$  are shown in figure \ref{fig:triple}.  Relative scaling of the A and BC contributions was achieved using our resolved NIRC2 photometry, while scaling between H and K bands was determined using our IRTF spectrum of the unresolved 2M0838$+$15 system.  The final spectra are sampled with two bins per resolution element at a native resolution of R$\sim$3800, but have been binned (using an error-weighted mean) by a factor of 7 to increase the signal to noise.  The $K$ band spectrum of component A has a very low signal to noise and we were unable to extract a clean spectrum free of systematic wiggles.    Thus we only show a rough SED for this component, in 0.03\,$\mu$m bins.
For reference we have overplotted the unresolved IRTF spectrum, and find it is reasonably well reproduced by the A+BC OSIRIS spectra.  Our OSIRIS spectroscopy confirms that both the A and BC components have T spectral types (discussed further in section \ref{sect:spt}).  Given the near equal luminosity and colors of the BC components, this result confirms that all 3 constituents of the 2M0838$+$15 system are T-dwarfs.
  
\section{2M0838$+$15 ABC: Discovery of a Visual Triple T-Dwarf System }
 \label{sect:triple}
 \subsection{Analysis of the NIRC2 images: binary system properties and component fluxes}
 \label{sect:mpfit}
\begin{figure*}
\epsscale{1}
\plotone{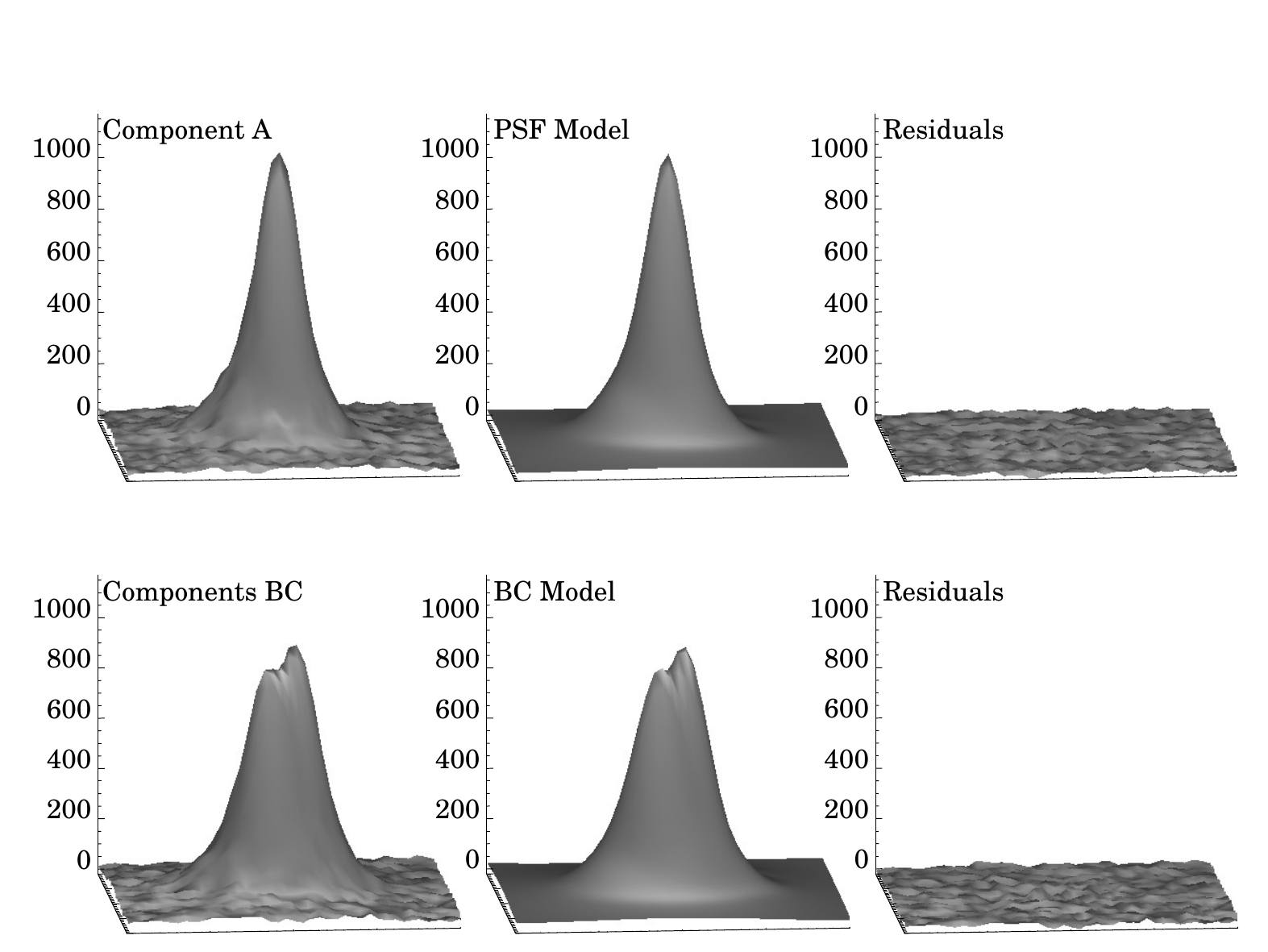}{
\caption{\redcol A series of $41\times41$ pixel surface plots demonstrating our PSF fitting of the 2M0838$+$15 BC system.  Component A is shown on top, and components BC on the bottom.  From left to right: the data, the model, and residuals.  \label{fig:psf}}}
\end{figure*}

{\redcol Binary parameters for the tight BC components were determined by fitting a double PSF model (using component A as a single PSF reference) to the data.  We modeled component A as a sum of 2D Gaussians, 

\begin{equation}
M_A(x,y) =\sum_{k=0}^N A_k e^{u_k}  + p_0 + p_1 x +p_2 y
\end{equation}

where  

\begin{equation}
\begin{array}{lll}
u_k & = & \left(\frac{(x-x_k)\cos{\theta_k}}{\sigma_{k,1}}-\frac{(y-y_k)\sin{\theta_k}}{\sigma_{k,1}}\right)^2 + \\  
& & \left(\frac{(x-x_k)\sin{\theta_k}}{\sigma_{k,2}} + \frac{(y-y_k)\cos{\theta_k}}{\sigma_{k,2}}\right)^2
\end{array}
\end{equation}

Above, $x$ and $y$ are pixel coordinates, $x_k$, $y_k$, $A_k$, $\sigma_{k,1}$, $\sigma_{k,2}$ are parameters describing the position, amplitude, and widths of the $k$th Gaussian along major and minor axes, $\theta_k$ is a rotation of the $k$th Gaussian with respect to the pixel coordinate system, and parameters  $p_0$, $p_1$, $p_2$ define a background plane.  The fit is constrained to disallow high frequency features by requiring that $\sigma_{k,1}$ and $\sigma_{k,2}$ $>$2.12 pixels (FWHM $>$ 5 pixels).  Our model of the single PSF then consists of $6N_g$+3 parameters where $N_g$ is the number of Gaussian components\footnote{We find that up to 7 components are needed to account for substructures in the PSF that are above the noise floor}.  For each image we fit our multiple Gaussian model to component A using Levenberg-Marquardt least-squares minimization as implemented in the IDL software  {\tt MPFIT} \citep{mpfit}.   The fit was performed within a 39 pixel box centered on source.  We fit models with $N_g$ ranging from 1 to 9 to each image, and selected a final value of $N_g$ that minimizes the Bayesian Information Criterion \citep[$BIC$;][]{liddle07}, where  here  $BIC=\chi^2 + (6N_g+3)\ln{N_{\rm pix}}$, where $N_{pix}$ is the number of data points.
Next, we fit a double version of our single PSF model to the BC components,
consisting of 9 parameters:  two specifying the pixel position of component B relative to component A, $x_B$ and $y_B$;  two specifying the angular separation and position angle of component C from component B, $\rho_{BC}$ and $\theta_{BC}$; two specifying the fluxes or amplitudes of the B and C components with respect to component A, $F_{B}/F_{A}$ and $F_{C}/F_{A}$; and three parameters defining a plane $c_0+ c_1 x +c_2 y$ to allow for a sloping background.  Fitting was performed in a 27 pixel box centered on components BC using {\tt MPFIT}.  We adopted the NIRC2 plate scale of 0.009963\arcsec pix$^{-1}$ and 0.13 deg offset in position angle found by \citet{ghez08} in order to convert pixel coordinates to angular separations and position angles.   The images were not corrected for distortion before fitting.  Based on distortion corrections provided by \citet{yelda10} the differential distortion at distances similar to the BC component separation is $\sim$0.2\, mas, and across our entire fitting box is $<1$\,mas.

We fit our model for the A and BC components to each individual exposure taken in the $J$, $H$ and $K_s$ bands.  An example of the data, model, and residuals for a single image is shown in figure \ref{fig:psf}.   In most cases subtraction of the best fitting model from the data leaves no significant residuals.  The best-fit parameters and their uncertainties returned by {\tt MPFIT} (with uncertainties scaled by the reduced $\chi^2$ value) 
are provided in table \ref{tab:mpfit}.  We obtain final estimates of parameters by taking a weighted mean of the results found for individual images, excluding the two $J$ band images that are not well resolved ($\rho_{\rm BC}< 0.5$ FWHM).  We note that for most parameters the image-to-image variance is much larger than the uncertainties inferred from {\tt MPFIT}.  Therefore, we choose to adopt the RMS of all measurements as an estimate of the overall uncertainty in order to account for these systematic differences.   We find a binary separation of $\rho_{BC}=50.2\pm0.5$\,mas and a position angle of $\theta_{\rm BC}=-6^{\circ}\pm2^{\circ}$.   For the A(BC) system parameters we determined the average $x-$ and $y-$pixel offsets between component A and the midpoint of the BC system in each image, corrected for distortion using the pixel offsets provided by \citet{yelda10}, and then multiplied by the plate scale.   Averaging results from all images we find a separation and position angle of $\rho_{\rm A(BC)}=549\pm 1$\,mas and $\theta_{\rm A(BC)}=18.8^{\circ}\pm0.1^{\circ}$. Relative fluxes of the A, B, and C components were determined in a similar fashion and then converted to relative magnitudes.  The resultant system and component properties are provided in tables \ref{tab:sys_prop} and \ref{tab:comp_prop} respectively.}

\subsection{Common Proper Motion}
On 2010 March 22 followup images of the 2M0838$+$15 system were obtained in the $H$ and $CH_4s$ (off-methane) filters using the NIRC2 narrow camera.   Observing conditions were partially cloudy, the image FWHM was large (0.12\arcsec), and  the BC components were not resolved.  From fitting gaussian PSFs to these data, relative positions and fluxes were determined between the A(BC) components.  We found the ($CH_4s-H$) color of components A and BC  to be indistinguishable within $\pm$0.05 mag (with uncertainties inferred from 6 and 5 dithered images in the $CH_4$s and $H$ bands respectively), implying that the A and BC components have identical spectral types within $\pm$1 subtypes according to the spectral type versus ($CH_4s-H$) relationship provided by \citet{liu08}.  This provided strong initial confirmation that the A(BC) components were both early-mid T-dwarfs.

The 73 day separation between the first and second epoch NIRC2 images allowed us to confirm the common proper motion of the A(BC) components.  Based on 2MASS, SDSS and WISE epochs \citet{aberasturi11} determined a proper motion of $\mu=0.13\pm0.06$\,mas\,yr$^{-1}$ for the unresolved 2M0838$+$15 system (see table \ref{tab:sys_prop}), which would amount to a linear motion of $25\pm12$\,mas  between observations.  We find the separation of the A(BC) components remains unchanged between 2010 Jan 08 ($548.6\pm1.2$\,mas) and 2010 March 22 ($549.0\pm1.6$\,mas) within a combined 2\,mas uncertainty.  This allows us to constrain the relative motions of the A(BC) components to within $\pm$10\,mas\,${\rm yr}^{-1}$, or a tenth of the system common proper motion.  Given their similar spectral types, proximity on the sky and shared proper motions within 10\,mas\,${\rm yr}^{-1}$, we conclude that  the 2M0838$+$15 ABC components are physically associated.

\subsection{System and Component Properties Inferred from Empirical Trends and Evolutionary Models}

\subsubsection{Magnitudes and Colors}
We used relative fluxes measured from our model fitting in section \ref{sect:mpfit}, combined with 2MASS magnitudes to determine the individual magnitudes of the components.  Since the system is not detected in the 2MASS $K_s$ band we derived a 2MASS $K_s$ magnitude by computing a synthetic $J-K_s$  color of 0.45$\pm$0.07\,mag from the SpeX spectrum \citep[e.g.][]{radigan12}, which implies a $K_s=16.20\pm0.20$.  The quoted uncertainty takes into account the relative component fluxes derived in section \ref{sect:mpfit}, the photometric error reported in the 2MASS catalog, and uncertainties in our synthetic SpeX photometry, adding contributions from various sources in quadrature.  Individual magnitudes and colors of the components are given in table \ref{tab:comp_prop}.  

Since we have determined relative component fluxes and magnitudes more precisely than system magnitudes reported in 2MASS, we include values and uncertainties for the differential magnitudes between components, $\Delta J$, $\Delta H$, and $\Delta K_s$ in table \ref{tab:comp_prop}.

\subsubsection{Spectral Types}
\label{sect:spt}
We determined spectral types for the A, B and C components using the resolved NIRC2 photometry and spectral templates of other field brown dwarfs from the SpeX Prism Library\footnote{Maintained by A. J. Burgasser, and located at http://pono.ucsd.edu/$\sim$adam/browndwarfs/spexprism/library.html} to decompose our unresolved 2M0838$+$15 SpeX spectrum into its individual ABC components, described in detail below.  We first decomposed the unresolved system into A and BC components.  {\redcol Next, we subtracted the best-fitting component A template from the unresolved system spectrum, and then decomposed the remaining BC contribution into individual B and C components. } 

\begin{figure*}
\begin{center}$
\begin{array}{cc}
\includegraphics[width=3in]{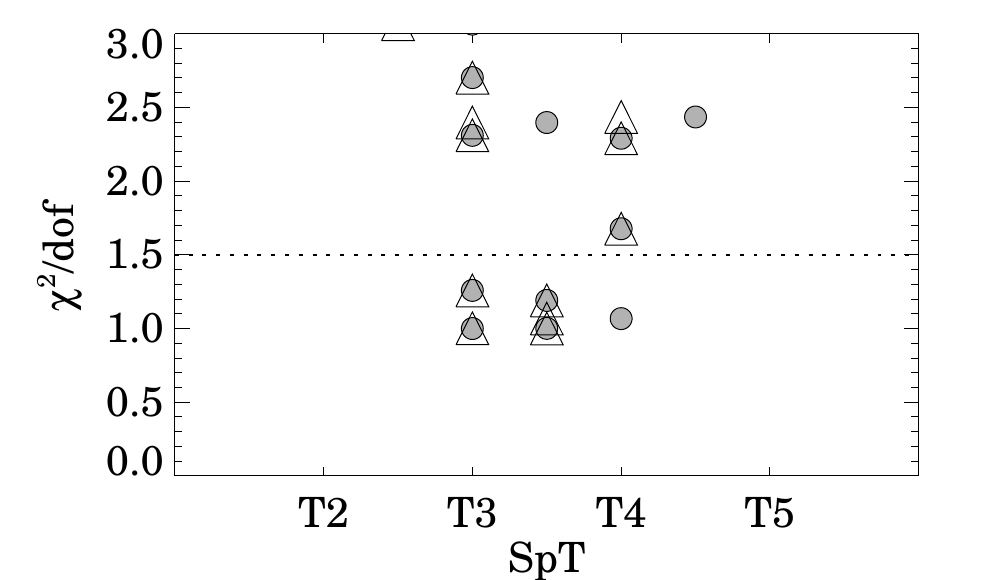} &
\includegraphics[width=3in]{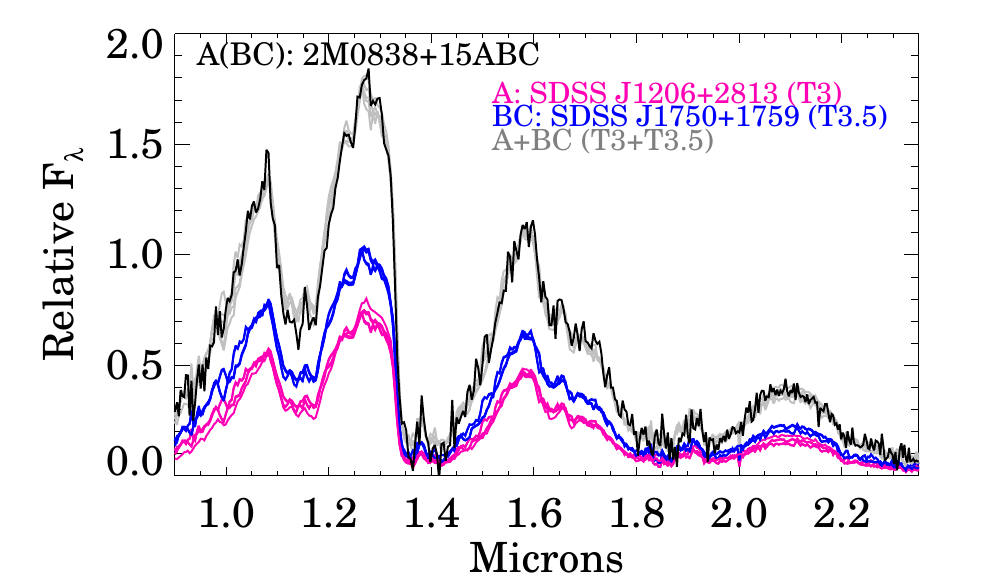} \\
\includegraphics[width=3in]{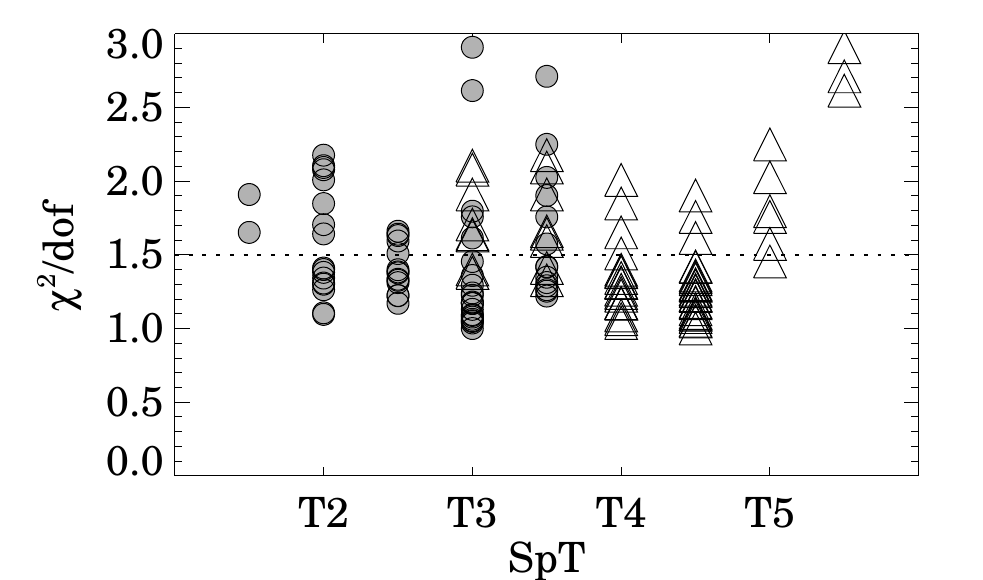} &
\includegraphics[width=3in]{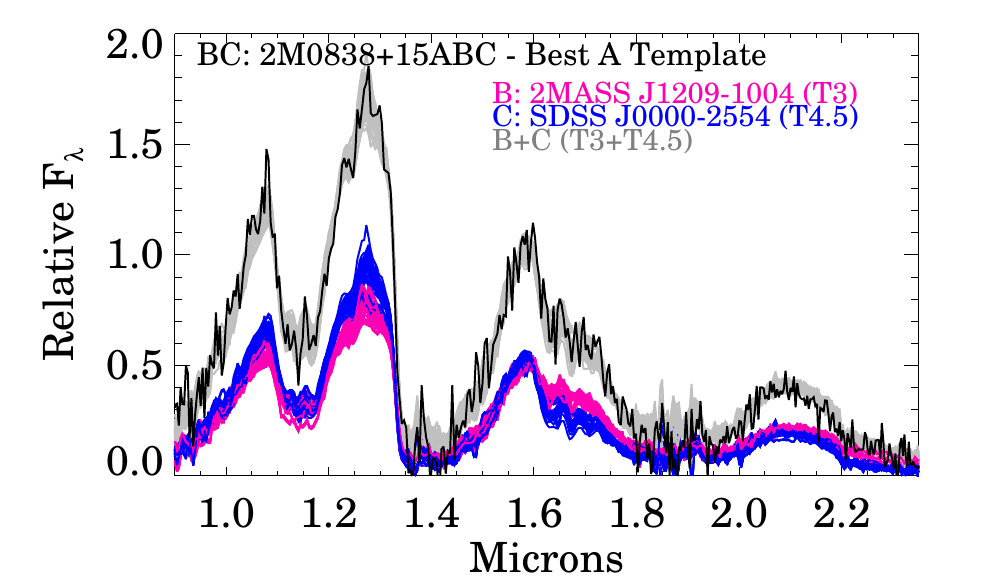} \\
\includegraphics[width=3in]{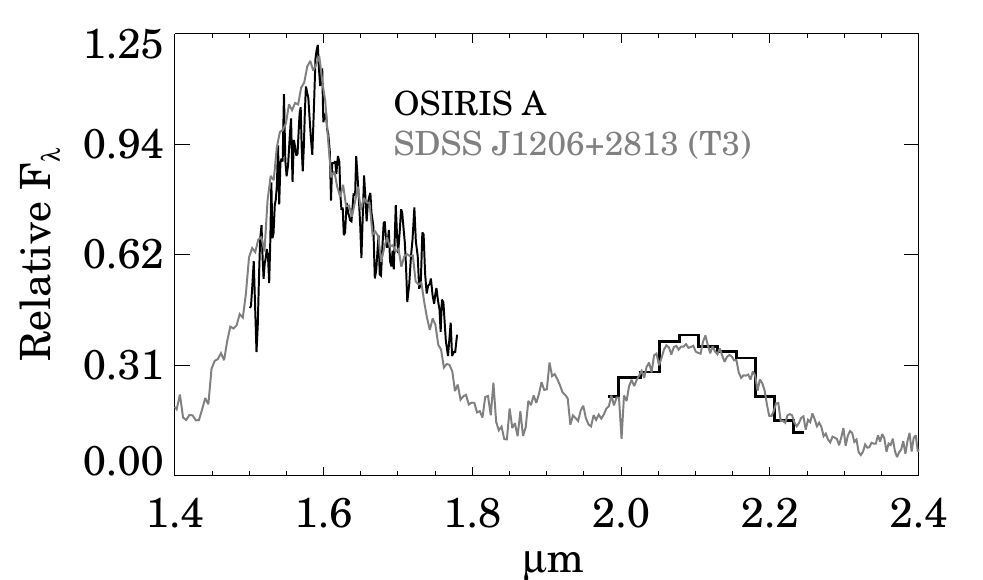} &
\includegraphics[width=3in]{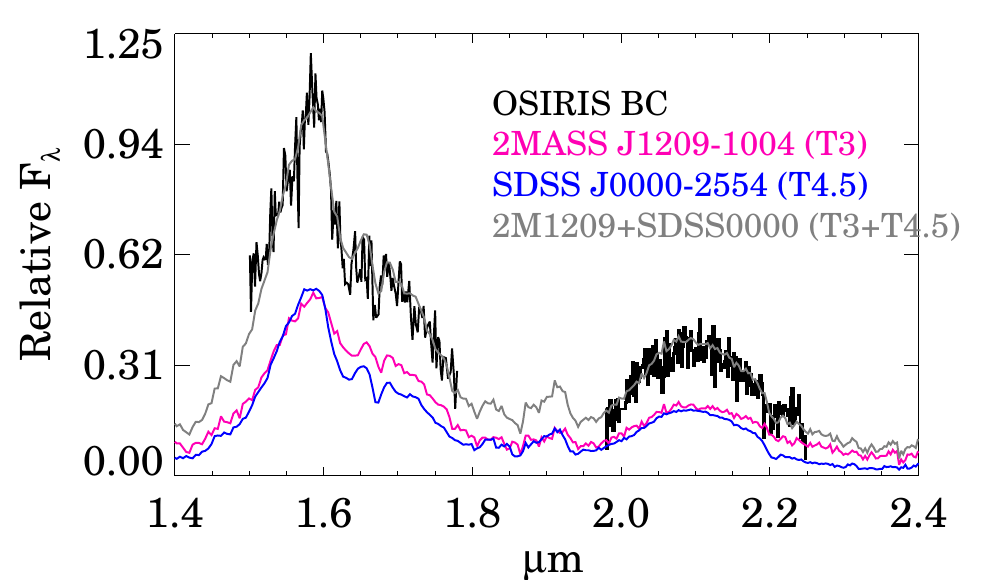}  
\end{array}$
\end{center}
\caption{ \redcol Top panel:  spectral decomposition of the A(BC) components using composite spectral templates from the SpeX prism library as described in section \ref{sect:spt}.  The reduced $\chi^2$ values for each spectral fit as a function of A (filled circle) and BC (open triangle) spectral types are shown in the left panel.   The composite templates with $\chi^2/\rm{dof} < 1.5$ (grey lines) are shown on the right, in comparison to the unresolved SpeX spectrum of 2M0838+1511 (black line).  Contributions from the individual A and BC templates are shown in cyan and pink respectively.  The best-fitting A and BC templates are indicated with pink and cyan labels respectively.  Note that each point on the left belongs to a unique {\em composite}  template, but that individual A and BC template spectra may each contribute to multiple composites.   Middle panel: spectral decomposition of the BC components, in a similar fashion to the A(BC) decomposition in the top panel.  {\em Bottom panel:}  the best fitting spectral templates for the A, B and C components, over plotted on the resolved OSIRIS spectra for components A and BC.  \label{fig:spec_a}}
\end{figure*}

{ To perform the decomposition, we first identified all templates with spectral types $>$T0 in the SpeX prism library sharing the same spectral resolution (R$\sim$120) as our 2M0838$+$15 ABC spectrum\footnote{\redcol We note that limiting the templates to spectral standards, we were unable to find a good match to the integrated light spectrum of 2M0838$+$15}.  Since the relative color between components is much better constrained than the system color we first identified all templates sharing the same $J-K_s$ color as component A within 1$\sigma$ uncertainties. For each A-template we then identified templates for component BC falling within the observed $\Delta(J-K_s)\pm \sigma_{\Delta(J-K_s)}$  and $\Delta(J-H)\pm \sigma_{\Delta(J-H)}$ between A and BC components.   The prospective A and BC templates were scaled to match the observed $H$-band fluxes of their respective components and added together to create a composite template.  The composite templates were then interpolated onto the data and scaled in order to minimize $\chi^2 = \sum{_i[({\rm data_i} - {\rm template_i})/{\rm error_i}]^2}$.  We scaled the error contribution in order to achieve $\chi^2 / {\rm dof} =1$ for the best fitting model.  Results are shown in figure \ref{fig:spec_a}.  Acceptable visual matches for the A(BC) components (deemed to occur for $\chi^2/{\rm dog}<1.5$) range from T3-T4 with the optimal match for component A being the T3 dwarf SDSS~J120602.51$+$281328.7 \citep{chiu06}, and the optimal match for component B being the T3.5 dwarf SDSSp~J175032.96$+$175903.9 \citep{geballe02,burgasser06}.  We therefore estimate spectral types of T3$\pm1$ and T3.5$\pm$1 to the A and BC components respectively.

{\redcol  The BC components were decomposed in a similar manner after subtracting the best fit for component A from the unresolved system spectrum.  We found best fitting templates with T3 \citep[2MASS J12095613-1004008;][]{burgasser06} and T4.5 \citep[SDSS~J000013.54$+$255418.6;][]{burgasser06}}
spectral types for the B and C components respectively, with reasonable visual matches encompassing T2-T4.5 and T3-T5 spectral types.  We therefore estimate spectral types of T3$\pm$1 and T4.5$\pm$1 for the unresolved  B and C components.  The results of our spectral decomposition are shown in figure \ref{fig:spec_a}, and provide a reasonable match to the resolved OSIRIS A and BC spectra.  }

The similar spectral types found for the A and BC components are consistent with (CH$_4$s - $H$) colors found in section \ref{sect:survey}, which indicate identical spectral types within $\pm$1.  In addition, the estimated A and BC spectral types agree well with the combined light spectral type of T3.5.  Due to the low SNR of the OSIRIS spectra, as well as sub-ideal sky subtraction and telluric correction we have not attempted to obtain spectral types by direct fits to these spectra, or from narrow band indices.
In the future, higher SNR spectroscopy, possibly resolving all three components should be attempted.

\subsubsection{Absolute Magnitudes and Distance}
Using the 2MASS $M_{K_s}$ vs SpT relation of \citet{dupuy12} and magnitudes and SpTs derived above for the ABC components, we calculated an absolute magnitude and distance modulus for the individual A, B and C components.   We determined a mean distance modulus of 3.5$\pm$0.5\,mag, which implies a distance of 49$\pm$12\,pc.  Distances inferred for the individual A, B and C components of 46, 54 and 47\,pc agree within uncertainties, yet reflect the similar spectral types despite different brightnesses for the A and B components.  Given uncertainties take into account uncertainties in the measured magnitudes and spectral types of the components, as well as intrinsic scatter of $\sim$0.46\,mag about the Dupuy \& Liu et al. relationship.

At a distance of 49\,pc and assuming the semi-major axes to be similar in length to their projections {\redcol (a reasonable assumption given correction factors of $a/\rho$=0.85-1.16 for visual VLM binaries in \citet{dupuy11b}), the system would have physical separations of 2.5$\pm$0.5\,AU (BC) and 27$\pm$5\,AU (A[BC]) respectively.}

\subsubsection{Relative Masses, Effective Temperatures and Surface Gravities}
Without an independent measurement of mass or age, evolutionary models are degenerate and prevent absolute physical properties from being inferred.  However, assuming the components of the ABC system to be coeval we can infer relative properties for a reasonable range of ages.  We estimated bolometric luminosities using the updated MKO-$K$ band bolometric corrections provided by \citet{liu10}, first converting 2MASS $K_s$ magnitudes to MKO magnitudes using the the corrections as a function of spectral type provided by \citet{stephens04}.   The differences in bolometric corrections between components is small, with a maximum difference of 0.013 mag between components B and C.  Thus relative bolometric luminosities inferred are dominated by the differences in $K$ band magnitudes of the components rather than their bolometric corrections.    This yielded bolometric luminosities of $\log{L_{A}/L_{\odot}}=-4.92\pm0.21$\,dex, $\Delta\log{L_{B-A}/L_{\odot}}=-0.13\pm0.03$\,dex, and $\Delta\log{L_{C-A}/L_{\odot}}=-0.23\pm0.04$\,dex, where we have expressed the luminosities for the B and C components as differences relative to component A. 

\begin{figure*}
\epsscale{0.7}
\plotone{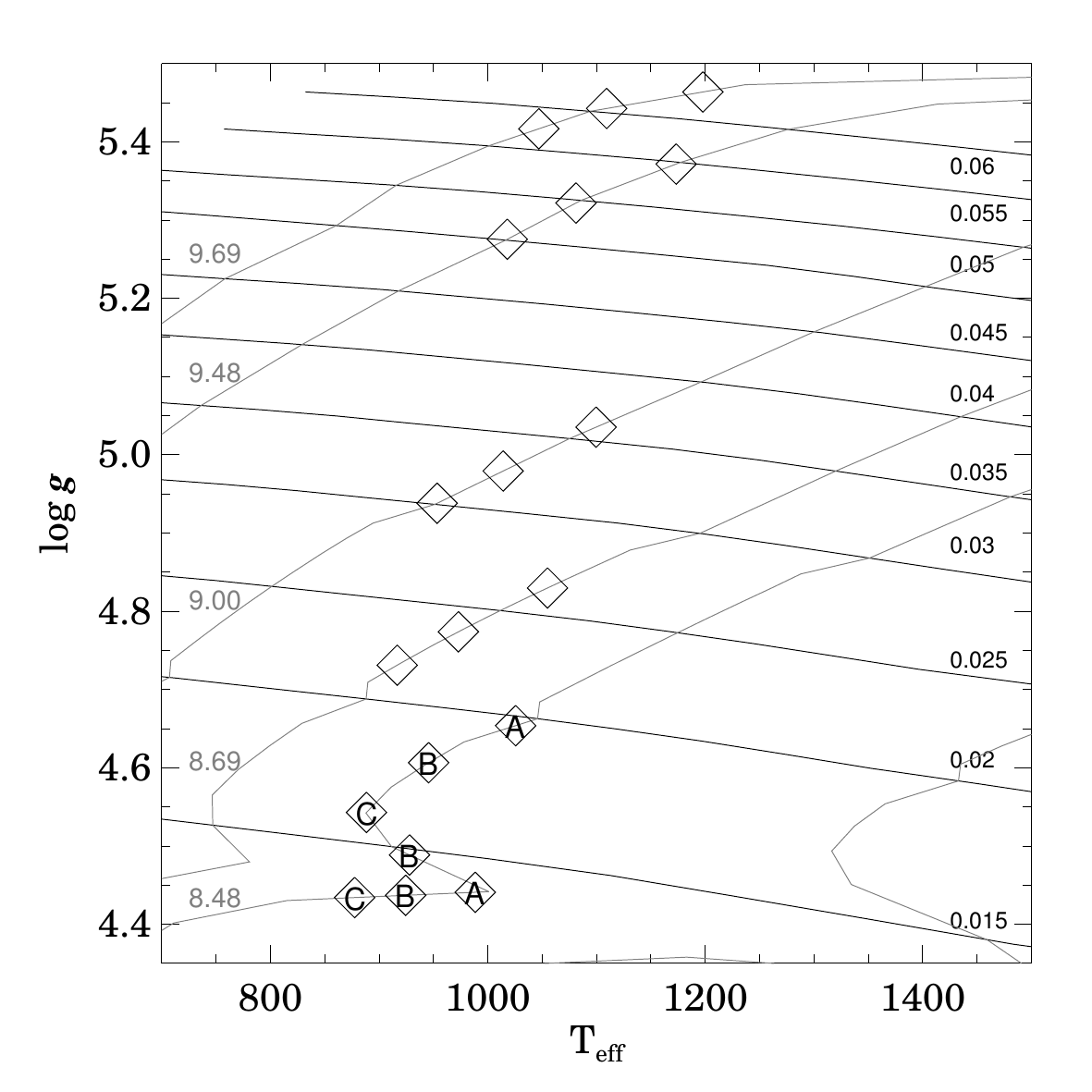}
\caption{\redcol An example of the age-mass degeneracies typical for BD systems, with surface gravities and effective temperatures for the 2M0838$+$15  ABC components inferred using empirical trends and evolutionary models, are plotted (from left to right) along a series of isochrones ranging from 300\,Myr to 5\,Gyr.  Isochrones (grey lines labeled with $\log{t {\rm [yr]}})$ and lines of constant mass (black lines labeled in units of $M_{\odot}$) using the evolutionary models of \citet{burrows97} are overplotted. Non-singular solutions for the 300\,Myr isochrone are shown with the components labelled by letter.  A dynamical mass for the BC system will constrain the system age, temperatures and surface gravities, and allow a direct comparison to model isochrones. \label{fig:iso}}
\end{figure*}

For a given system age these luminosity estimates can be converted to masses, effective temperatures, radii, and surface gravities via evolutionary models.  We used the  evolutionary models of \citet{burrows97} to infer masses, effective temperatures, and surface gravities at ages of 3~Gyr (typical for field brown dwarfs) and 300 Myr.  For a 3~Gyr old system we find masses of 60, 55, and 50 $\times 10^{-3} M_{\odot}$ with a systematic uncertainty of $\sim$15\% and relative uncertainties of $\sim$3\%.  For an age of 300~Myr we find masses of 20, 18 and 16 $\times 10^{-3} M_{\odot}$ with a systematic uncertainty of $\sim$25\% and relative uncertainties of $\sim$5\% (we note that there are degenerate solutions for this age, which yield component masses as low as $\sim$$0.13-0.14\,M_{\odot}$; see figure \ref{fig:iso}).  The largest source of uncertainty factoring into these calculations comes from the absolute magnitude determination.

In both cases the mass ratio of the BC components, $q_{\rm BC}$ is close to unity (0.92 for 3\, Gyr and 0.89 for 300\, Myr), which is typical for brown dwarf binaries which are found to have a mass ratio distribution peaking strongly at unity \citep[e.g.][]{burgasser07_bin,allen07,liu10}.  While the A(BC) mass ratio of $\sim$0.57 would be atypically low compared to the majority of BD binaries, near-equal mass A, B, and C components for VLM triple systems are common \citep[e.g][]{burgasser12}.

In order to estimate a range in surface gravities and temperatures spanned by the ABC components we
have plotted inferred effective temperatures and surface gravities along a series of isochrones ranging from 300\,Myr to 5\,Gyr, shown in figure \ref{fig:iso}.  
For ages of 0.5-5 Gyr  $d(\log{g})/d(T_{\rm eff}$) is approximately constant along isochrones, with the ABC components spanning a fairly narrow range in temperature ($\sim$150\,K), and surface gravity ($\sim$0.1 dex) irrespective of age or mass.   

Given the inferred system masses and a semi-major axis of of 2.5\,AU, the BC system would have a period ranging from $\sim$12\,yr (3\, Gyr) to 21\,yr (300\,Myr).  Thus a dynamical mass measurement of the BC components may be possible on a relatively short timescale, which should greatly constrain the system's position in figure \ref{fig:iso}.  In addition, efforts to obtain a parallax are ongoing, which will provide improved constraints on the system's absolute magnitude and bolometric luminosity.

\section{Search for companions and sensitivity limits}
\label{sect:survey}

Here we present our search for companions around the entire sample of 8 L/T transition objects observed with NIRC2 (section \ref{sect:nirc_obs}).  

Each reduced image was carefully visually examined for companions.  Only a single object, 2M0838$+$15, was resolved into a multiple system.    Our sensitivity to companions was determined via simulations.  For a given binary separation, $\rho$, and contrast, $\Delta K$, we constructed 100 simulated binary pairs using a cutout of the observed target as a single PSF model, with randomized position angles.  We then attempted to recover companions from the simulated images in the following way:

\begin{enumerate}
\item We first subtracted off a 31-pixel median filter of the input image in order to remove any sloping background and then iteratively measured the RMS background noise (i.e. in regions without a source) of the simulated image, $\sigma$.
\item We searched for point sources in the image using the IDL adapted version of the DAOPHOT \verb|find| algorithm, setting a 5$\sigma$ detection threshold.
\item For each of the candidates identified by \verb|find| we measured the RMS noise in a 1 FWHM  wide annulus with a mid-radius corresponding to the distance to the candidate source.  The candidate source itself was masked using a circular region with 0.75 FWHM radius.  Candidates at separations greater than 350\,mas and with peak fluxes less than 5 times the noise within the annulus were discarded.  For candidates at separations less than 350\,mas we enforced a slightly higher detection threshold of 7 times the noise, due to the presence of high frequency structure in the AO PSF in this region.  
\item A further search for close (blended) companions inside 200\,mas  was conducted.  Low frequency features were subtracted from the simulated image using a median-filtered image (wherein each pixel is replaced by the median of surrounding pixels in a box size of 1$\times$1 FWHM rounded up to the nearest odd-integer number of pixels).  We then repeated our search for additional peaks using \verb|find|.  Sources with peak fluxes $<$0.15 times that of the central source were discarded.  This detection limit corresponds to roughly three times the size of the largest residuals found when subtracting a gaussian PSF from our data.  
\end{enumerate}

\begin{figure*}
\epsscale{1.1}
\plotone{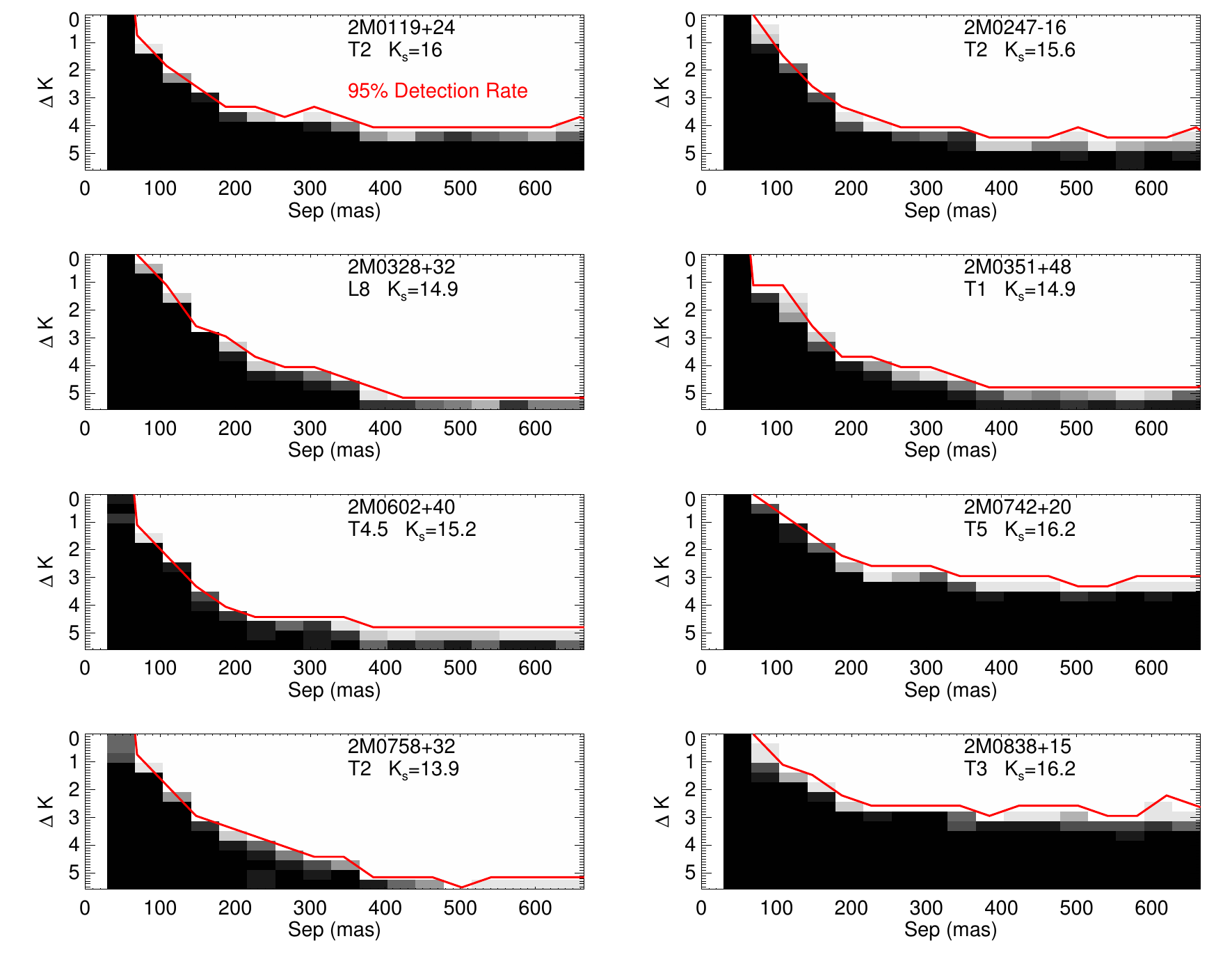}
\caption{$K$-band detection limits as a function of binary separation and  contrast, $\Delta K$.  Red lines indicate the 95\% recovery rate of simulated companions.  The shading depicts the recovery rate for any separation and contrast, linearly varying from a 0\% percent (black) to 100\% (white) \label{fig:sens_k}}
\end{figure*}

The above criteria were verified by visual inspection and consistently picked out bonafide companions in our simulated datasets, while rejecting other point-source-like structures in the AO PSFs.    We note that since the PSF is significantly different for each target (dependent on the relative position and brightness of the TT star used) we were unable to build a model of a single PSF to subtract from those of our targets.    When applied to the data, the above criteria successfully detect the close 2M0838$+$15BC pair in the science images wherein $\rho_{{\rm BC}} >0.5$ FWHM, while assigning non-detections to the other sources in our sample.  Our sensitivity to companions at a given separation is given by the recovery rate of simulated companions, and is shown for each source in figure \ref{fig:sens_k}.

Although varying from target to target we are sensitive to  approximately 95\% of companions with $\Delta K<1$  at innermost separations of 60-120\, mas,  and to companions with contrasts of $\Delta K>$~2.5-4 at separations $>$200\,mas.  The tight BC components of 2M0838$+$15 fall inside the minimum separation  where companions are routinely detectable.  We easily resolve the BC pair due to the fact that the binary axis runs approximately perpendicular to the PSF elongation in the direction of the TT star.  In other words, sensitivity is a function of position angle for elongated PSFs, and the 95\% recovery rates reported reflect the least sensitive position angles.

\subsection{The binary fraction of our sample \label{sect:fbin}}
In our sample we resolved 1 of 8 targets into a multiple system (we do not count 2M0838$+$15 BC as an additional target).  The probability of observing $n$ multiples in our sample is given by the binomial distribution, $P(n | N=8, \nu_{obs})$, where $\nu_{obs}$ is the observed binary frequency (uncorrected for observational biases), $n$ is the number of binaries observed and $N$ is the sample size.  A Beta distribution for $\nu_{obs}$ is obtained by using Bayes Law to infer  $P(\nu | n=1, N=8) \propto P(n=1 | N=8, \nu_{obs}) P(\nu_{obs})$, where we have assumed a flat (most ignorant) prior probability distribution of $P(\nu_{obs})=1$.  From the posterior distribution we derive an observed multiple frequency of 12.5$^{+13.4}_{-8.2}$\% for our sample.
The quoted uncertainties correspond to the 68\% credible interval of the distribution of frequencies.  For a non-symmetric distribution there are many ways to construct a credible interval about the maximum likelihood.  Here we have constructed a ``shortest'' credible interval $[a,b]$, such that $P(a)=P(b)$, which is more informative than an equal-tail interval.  Within the uncertainties, this result is consistent with those from other magnitude-limited studies which find uncorrected binary fractions of $\sim$17-20\% \citep[e.g.,][]{bouy03,gizis03,burgasser03, burgasser06}.

\section{Binary statistics in the L/T transition}
\label{sect:stats}
\subsection{The L9-T4 binary frequency}
{\redcol The L/T transition is roughly the regime over which condensate clouds disappear as a major opacity source in brown dwarf photospheres.  Here we consider the L/T transition to encompass L8-T5 spectral types, which is the range associated with $J$ band brightening.  We have chosen not to include the ends of this branch (L8 and T5 bins) in our analysis so as not to contaminate the sample with objects of ambiguous membership.  There is a long standing question as to whether the binary fraction might be larger inside the transition.  \citet{burgasser07_ltbin} demonstrated with simulations that a flattening of the luminosity function for single objects within the L/T transition, will result in a sparsity of objects in these spectral type bins.  
Combined with no such flattening of the unresolved binary luminosity function at similar spectral types, this results in an enhanced binary fraction.  
However, evolutionary models used by \citet{burgasser07_ltbin} did not account for cloud evolution, and more recent evolutionary models including evolving clouds \citep{saumon08}, predict a pileup rather than deficit of objects at the cloudy/clear transition.  Thus, observations of the L/T transition binary fraction can provide a useful test for evolutionary models.  Additionally, the L/T transition is the spectral type range over which variability may be expected due to heterogeneous cloud coverage \citep[e.g.][]{burgasser02_lt,radigan12}, and the level of binary contamination in this regime has important consequences for variability surveys \citep[e.g.][]{clarke08,radigan11}.}

Although small, our sample increases the number of objects observed with L9-T4 spectral types by 40\%.  If taken together with the L/T transition sample observed by \citet{goldman08}, these more recent data nearly double the number of objects surveyed for multiplicity in this regime.  Here we combine our data with the L/T transition survey of \citet{goldman08}, T-dwarf surveys of \citet{burgasser03, burgasser06}, and the combined L and T samples of \citet{bouy03} and \citet{reid06,reid08} to examine the binary frequency in the L/T transition.  

\subsubsection{The combined statistical sample}
Although the surveys considered are a heterogeneous group, they are all sensitive to minimum angular separations of approximately $\gtrsim0.05-0.1$\arcsec.  In addition, ultra cool dwarf binaries tend to have flux ratios well above typical detection limits (except at small angular separations), implying that differing sensitivities to faint companions among surveys does not strongly impact binary statistics \citep[e.g.][]{burgasser07_bin,allen07}.  To achieve consistency each target was cross-correlated with the DwarfArchives database of known L and T dwarfs to determine a homogeneous set of NIR spectral types and colors.  Names, spectral types and $J-K_s$ colors of the combined sample are provided in table \ref{tab:sample}.   

In total the combined sample consists of 19 objects including 2M0838$+$15 with NIR spectral types of L9-T4 (inclusive).  However, as the only known higher order multiple in the sample, 2M0838$+$15 is significantly further away ($d\sim$50\,pc) than other objects of the sample ($d<35$\,pc) and likely biases the result.  Rather than attempting a completes correction for higher order multiples we apply a 35\,pc distance cut, which effectively removes 2M0838$+$15 from the targets considered, yielding a final sample size of 18.

Based on 4/18 detections in the L9-T4 sample, we computed $P(\nu_{obs} | n, N) \propto P(n | N, \nu_{obs})$ as in section \ref{sect:fbin}, obtaining an observed binary fraction for the combined sample of $\nu_{obs}=22_{-8}^{+10}$\%.  This is on the high end, but comparable to other reported visual binary fractions for L and T dwarfs (subject to similar selection effects and observational constraints, and uncorrected for biases) previously discussed in section \ref{sect:fbin} ($\sim$17-20\%).

\subsubsection{Correcting for observational biases}
In magnitude limited samples a Malmquist bias leads to binaries being sampled in a larger volume than singles.  
Here we follow the example of \citet{burgasser03,burgasser06} who provide an expression for the real binary frequency, $\nu$, in terms of the observed frequency, $\nu_{\rm obs}$:

\begin{equation}
\nu = \frac{\nu_{obs}}{\alpha (1-\nu_{obs})+\nu_{obs}}
\label{eq:bin}
\end{equation}

where $\alpha$ is the ratio of volume searched for binaries to that of the volume searched for single objects given by

\begin{equation}
\alpha = \frac{\int_{0}^{1} (1+\rho)^{3/2}f(\rho)d\rho}{\int_{0}^{1} f(\rho)d\rho}  
\label{eq:alpha}
\end{equation}

where $\rho=f_2/f_1$ is the flux ratio of components, $f(\rho)$ is the distribution of flux ratios.  
The limiting cases where $f(\rho)$ is flat and 100\% peaked at unity yield values of $\alpha=1.86$ and $\alpha=2.82$ respectively, and typically an intermediate value is used.  However, if we examine the distribution of flux ratios for all known ultracool binaries (see Appendix A) we find a broad distribution that peaks at $\sim0.4-0.5$ with a negative skew, such that the mean flux ratio is slightly larger than this.  Integrating over this distribution we obtain $\alpha=1.87$, which is equivalent to the case where $f(\rho)$ is flat.   As a check, the binaries in our sample have flux ratios (in their survey filter, see table \ref{tab:sample}) ranging from 0.39-0.46 which yield an average value of $(1+\rho)^{3/2}=(d_{\rm bin}/d_{\rm single})^3$ of 1.7. As a compromise between the two values, we adopt a value of $\alpha=1.8$ for our sample.

To correct for our greater sensitivity to binaries over single objects, we performed a change in variables from $\nu_{obs}\rightarrow \nu$:

\begin{equation}
P(\nu) = P(\nu_{obs})  \left \lvert \frac{~d \nu_{obs}}{d \nu} \right \rvert   
\label{eq:nu}
\end{equation}

 where a relationship between $\nu$ and $\nu_{obs}$ is given in equation \ref{eq:bin}.    
 
 From the resulting distribution we determined the most probable binary frequency and uncertainties corresponding to a 68\% shortest credible interval.  This yielded a resolved L9-T4 binary fraction of $13^{+7}_{-6}$\% (at projected separations $\gtrsim 1-2.5$\,AU).    It is important to note that this is only the visual binary frequency for the stated detection limits and ignores the small-separation wing of the semi-major axis distribution.  By some estimates spectroscopic binaries could be as numerous as resolved systems \citep{maxted05,joergens08}, increasing the total binary fraction by  a factor of 2.  
 
 For comparison, the T-dwarf surveys of \citet{burgasser03} and \citet{burgasser06} found bias corrected binary fractions of $9^{+15}_{-4}$\% and $12^{+7}_{-4}$\% (but used a slightly larger value of $\alpha=2.16$), while the late-M and L-dwarf surveys of \citet{bouy03} and \citet{reid08} found binary fractions of $\sim$10\% and $12^{+5}_{-3}$\% respectively (the latter survey was volume-limited). While we find a slightly higher binary frequency from L9-T4 spectral types, it remains comparable to those reported by other surveys at the 1$\sigma$ level. 
 
  It is nonetheless interesting to note that if we select the L9-T4 sample based on {\em primary} spectral types rather than unresolved system types the observed binary fraction drops to 2/16 (=12.5\%), which translates into a bias-corrected frequency of $6^{+6}_{-4}$\% or about half of the unresolved fraction (resolved spectral types of binaries for the surveys considered here are provided in table \ref{tab:fr}).  Thus even though we don't find a statistically significant increase in binary fraction for L9-T4 spectral types relative to other L and T dwarf samples, there is some evidence of a systematic increase in binary frequency between unresolved and primary spectral-type-selected samples by a factor of $\sim$2.  This is consistent with the population synthesis of \citet{burgasser07_ltbin}, which show that the binary frequency approximately doubles for unresolved L9-T4 dwarfs given a flat input distribution as a function of primary spectral type.  In addition, we cannot rule out an L9-T4 binary frequency as high as 21\%  (the upper limit of our 68\% credible interval), and it therefore remains possible that the unresolved binary frequency is much higher in the L/T transition.  
  Clearly, a larger sample of L/T transition objects is needed to make progress. 

\section{Discussion and Conclusions}
\label{sect:discussion}
\subsection{2M0838$+$15ABC: A benchmark triple of early T-dwarfs}
We have presented resolved imaging and spectroscopy of the first triple T-dwarf system, 2M0838$+$15 ABC.

With a dynamical mass measurement possible for the BC components,  2M0838$+$15 ABC will serve as a benchmark system in the poorly contained L/T transition regime.  Even without a dynamical mass, this system represents the largest homogeneous sample of early T-dwarfs to date, whose relative colors and spectral types can be used to test evolutionary models along a single isochrone.

\begin{figure*}
\begin{center}$
\begin{array}{cc}
\includegraphics[width=3in]{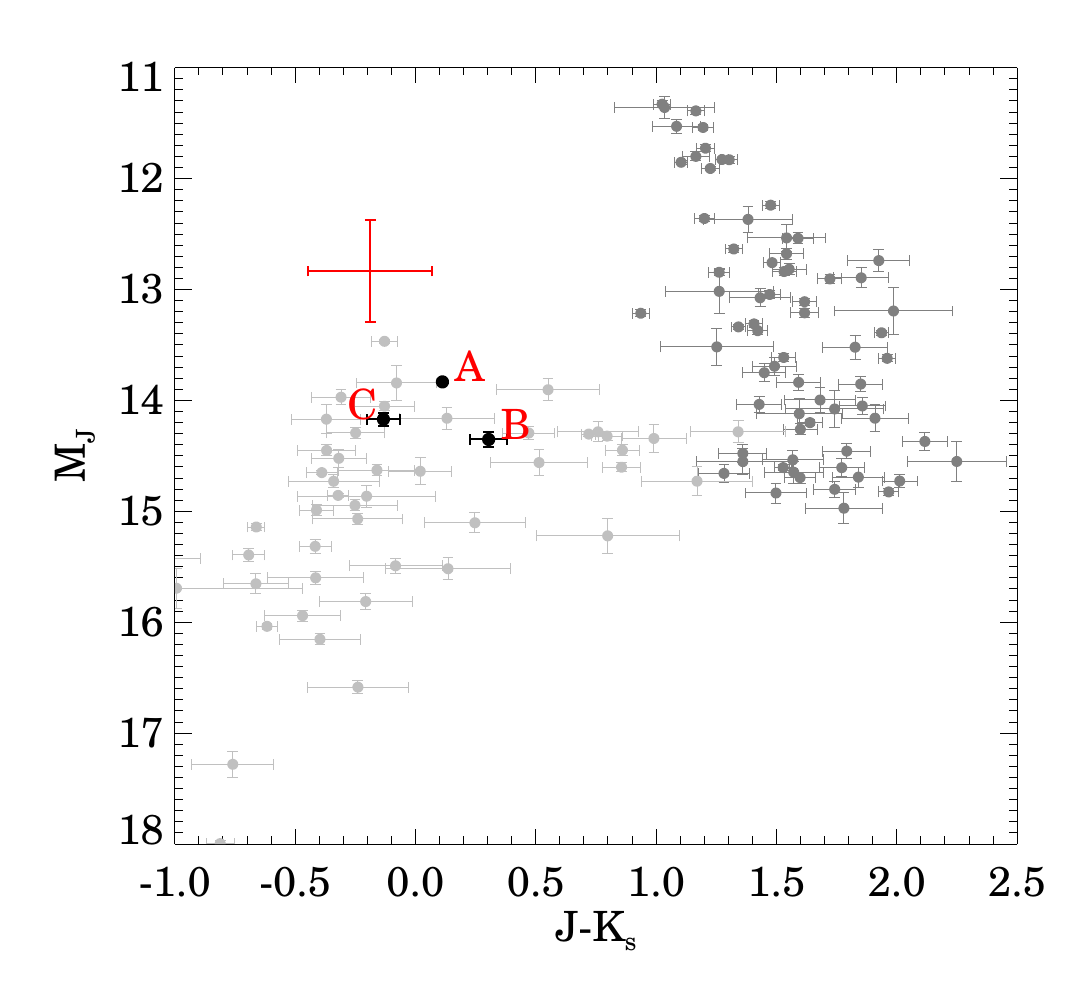} &
\includegraphics[width=3in]{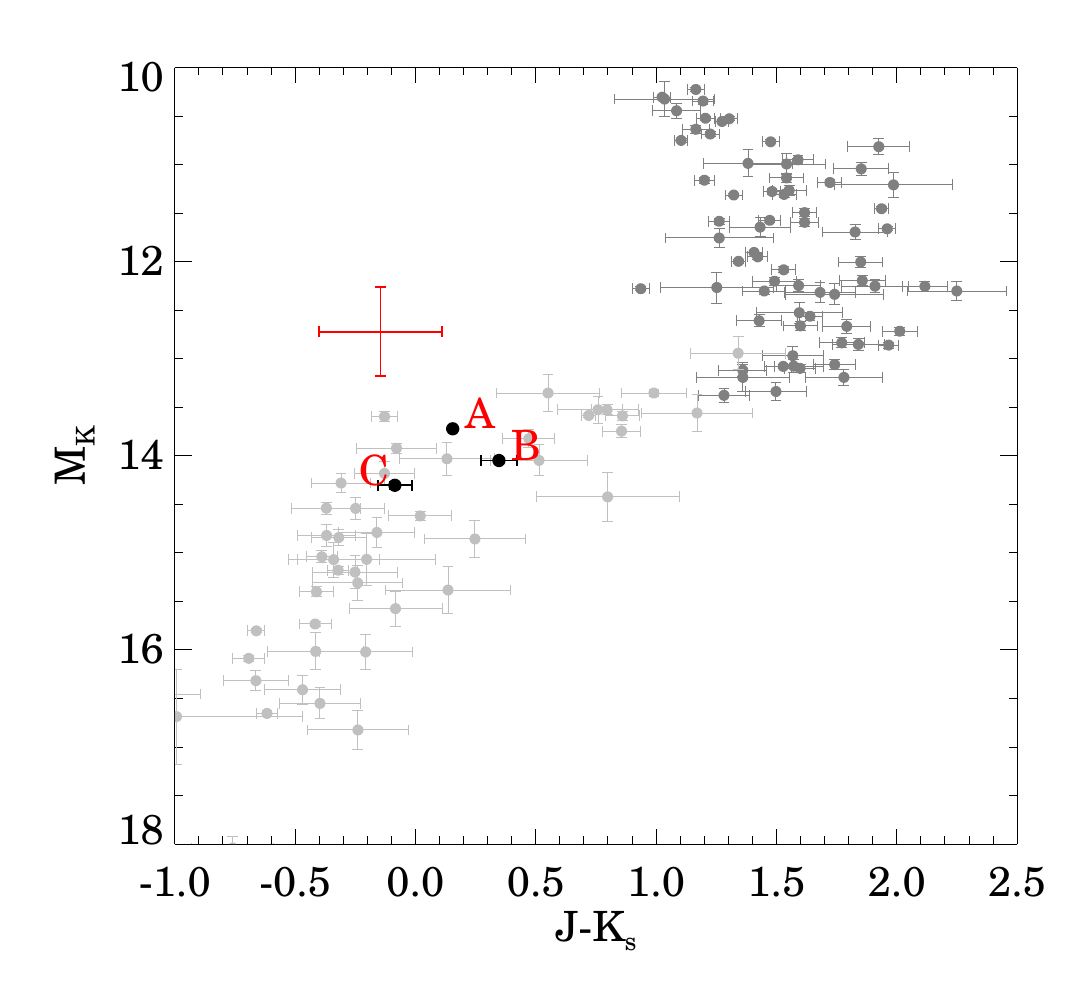}
\end{array}$
\end{center}
\caption{Color magnitude diagrams showing the {\em relative} positions of the 2M0838$+$15 A, B and C components inferred from resolved photometry of the system.  Note that the absolute position along the vertical $M_J$ and $M_K$ axes of the 2M0838$+$15 system is not independently constrained. In each figure, a red error bar shows the systematic uncertainty in absolute magnitudes and colors for the 2M0838$+$15 system, while black error bars show relative errors of the B and C component magnitudes and colors with respect to component A. \label{fig:jk}}
\end{figure*}

The relative positions of the A, B and C components on a color-magnitude diagram are shown in figure \ref{fig:jk}.  We find that component A, which is brightest in all three bandpasses, has a $J-K_s$ color intermediate to the fainter B and C components.  This could reflect a sensitive dependence of the L/T transition effective temperature on surface gravity, with the slightly higher mass/gravity component A evolving across the transition at a systematically higher temperature and luminosity than the BC components.  However, from figure \ref{fig:iso} we find that this difference in surface gravity can be no more than $\sim$0.1\, dex for ages greater 300\,Myr.

Alternatively, different condensate cloud properties could explain differences in brightness between components A and B without the need to invoke differences in surface gravity.  Observations of variable brown dwarfs have demonstrated that changes in cloud coverage at constant effective temperature, surface gravity, and empirical spectral type can lead to brightness variations as high as $\sim$26\% in the $J$ band and $\sim$15\% in the $K_s$ band \citep{radigan12}.   Thus rather than being more massive, it is possible that component A simply has thiner clouds and/or lower fractional coverage than components B and C. 

Assuming for now that 2M0838$+$15 does not vary in brightness, we explore the possibility that systematic differences in cloud properties between the A and BC components could be due to viewing geometry.   For instance, if the tight BC components have spin axes that are aligned with each other but not with the wider component A, the difference in relative cloud properties may reflect pole-on versus edge-on orientations of banded clouds.  There is some evidence both observationally \citep{hale94,monin06} and in hydrodynamic radiative simulations of star formation \citep{bate12} that close binary components (separation $\lesssim$ 30\,AU) have preferentially aligned spins on account of dissipative interactions during the formation process, while wider binaries and members of triple systems are more frequently misaligned.  Given that all field brown dwarf binaries have small separations ($\lesssim$10-15\,AU) it is possible that the majority have aligned spins and hence correlated cloud coverage, while higher order multiples such as 2M0838$+$15 ABC are more likely to have a severely misaligned component from three-body dynamics.   However, recent observations by \citet{konopacky12} of 11 VLM binaries cast some doubt on this hypothesis.   These authors find nearly half of their sample (5/11) have highly different $v\sin{i}$ (all with separations $<$ 5\,AU) which may indicate frequent spin mis-alignment in close VLM binaries (although this could also reflect intrinsically different rotation rates). 

\subsection{Formation and Dynamical Stability of 2M0838$+$15}
\label{sect:stab}
The 2M0838$+$15 ABC system stands out as the only known T-dwarf triple system to date, and furthermore the only known brown dwarf triple system where all three components have been directly and conclusively detected. There are only two other examples of possible or probable brown dwarf triple systems in the literature: Kelu-1 \citep{ruiz97,liu05,stumpf08} and DENIS-P J020529.0-115925 \citep{delfosse97,koerner99,bouy05}, although in former cases the tertiary component remains unresolved and in both cases flux ratios and separations cannot be accurately determined.  There are a handful of VLM triples (7 known or suspected to date) discussed in \citet{burgasser12}, and of these only two prove to be good analogs to 2M0838$+$15 ABC with near-equal component masses and relatively low ratios of outer-to-inner separations:   (i)LP 714-37,  an M5.5/(M8/M8.5) triple with similar component masses of 0.11, 0.09, and 0.08 $M_{\odot}$, and outer/inner separations of 33\,AU/7\,AU; (ii) LHS 1070 ABC, a (M5/M8.5/M9) triple with component masses of 0.12, 0.08, and 0.08 $M_{\odot}$ and separations of 12\,AU and 3.6\,AU \citep{leinert01,seifahrt08}.  The other known VLM triples have much larger inner-to-outer separation ratios ($\gtrsim 100$), which implies a  multimodal distribution of inner-to-outer period ratios as is seen for higher mass triples \citep[e.g.][]{tokovinin08}.

Hydrodynamic simulations of fragmentation and subsequent evolution in a gas-rich environment \citep[e.g.][]{bate02_closebin,bate09,bate12} can form triples at separations smaller than the fragmentation scale ($\lesssim$100-1000\,AU) and with preference for equal masses as a result of dissipative interactions with disks, and accretion.  These simulations successfully produce VLM and brown dwarf triples with outer-to-inner separation ratios of $\sim$10-100.  The role, if any, of subsequent dynamical interactions within a gas-free cluster is unclear.  Simulations of gravitational interactions between small-$N$ cluster members by \citet{sterzik03,delgado04} can infrequently produce BD triples ($\sim$0.2\% of all triple systems), but this mechanism does not reproduce observed properties of binaries \citep[it leaves larger populations of single stars and close binaries than are observed;][]{goodwin05}, or higher mass triples \citep[observed triples have broader period distributions, larger outer-to-inner period ratios, an mass ratios closer to unity than those formed in simulations;][]{tokovinin08}, and therefore is unlikely a major determinant of stellar multiplicity properties in general.  Nonetheless, there is nothing that rules out this scenario for the 2M0838$+$15 ABC system in particular, and the outer-to-inner period ratio of $\sim$10 is similar to those produced from dynamical decay simulations.  Furthermore, there is debate as to whether dynamical ejection \citep[e.g.][]{reipurth01,bate02_bd} of BD systems from the surrounding gas reservoir, while accretion is still ongoing, may be required to halt further growth and prevent proto-brown dwarfs from reaching stellar masses.   
 
\begin{figure*}
\epsscale{0.7}
\plotone{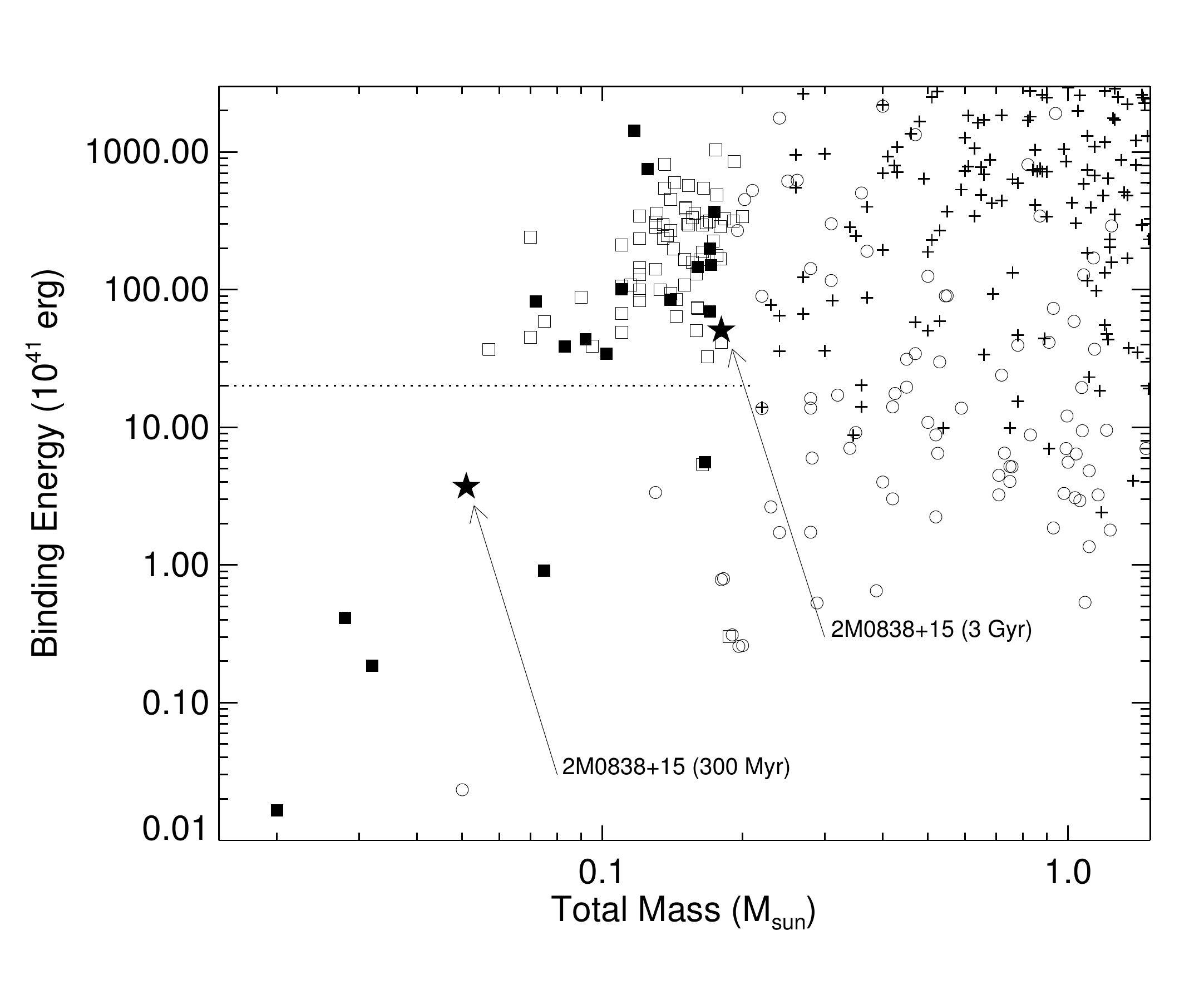} 
\caption{Binding energy as a function of total mass for stellar and VLM binary systems.  The 2M0838$+$15 A(BC) system is plotted using a 5-point star.  A dotted horizontal line makes the minimum binding energy cutoff of $\sim 20\times10^{41}$\,erg observed for the majority of VLM binaries.  Most of the other VLM binaries plotted (squares) are taken from the VLM Binary Archive, which is a compilation of data from 144 unique publications which can be accessed at \tt{vlmbinaries.org}.  {\rm Additional VLM systems, stellar binaries (crosses) and stellar primaries with VLM secondaries (open circles) are taken from \citet{close90,fischer92,tokovinin97,reid97a,reid97b,reid01,caballero06,daemgen07,kraus07,radigan08,luhman09,faherty10}\label{fig:sf}.  Young/low-gravity systems are marked by filled circles and squares.}}
\end{figure*}

For an age of 3\,Gyr we find an approximate binding energy of $\sim$20$\times 10^{41}$\, erg, satisfying the minimum binding energy typically found for VLM binaries \citep[e.g.][]{close03,burgasser07_bin}.  If this binding energy cut-off is the result of dynamical ejection then the 2M0838$+$15 ABC system is likely to have survived such an event.  
Alternatively, for an age of 300\,Myr the system would have a binding energy of $\sim$1.6$\times10^{41}$\,erg, well below the empirical minimum, and join only a handful of similarly weakly bound VLM and BD systems (see figure \ref{fig:sf}).  This observation may favor an older age for the 2M0838$+$15 ABC.  

With an inner-to-outer ratio of projected separations of $\sim$10, the 2M0838$+$15 ABC system currently satisfies the stability criterion suggested by \citet{eggleton95} of $Y_0>6.7$,  where $Y_0$ denotes the ratio of inner binary apastron separation, versus the outer binary periastron separation.   Due to projection effects the actual inner to outer separation ratios could be even larger than measured.  Assuming that the outer to inner ratio of semi-major axes is close to the observed ratio of projected separations  ($\sim$10) implies a period ratio of $\sim$30 for the inner and outer orbits.   In this case the empirical stability criterion of \citet{tokovinin04}, $P_{out}(1-e_{out})^3/P_{in}>5$, is satisfied for outer eccentricities of $e_{out}\lesssim0.45$.
Thus current observations are consistent with the longterm stability of the 2M0832+15ABC system.  

\subsection{Binarity at the L/T transition}
We have found a late-L/T transition binary frequency (L9-T4 spectral types, at observed separations $\gtrsim0.05-0.07$\arcsec)  of $13^{+7}_{-6}$\%.  This is similar to reported frequencies outside of the transition \citep[$\sim$9-12\%;]{burgasser03,burgasser06,reid08,bouy03}.  This preliminary result provides an optimistic outlook for studies of L/T transition brown dwarfs: it suggests that this sample is not significantly contaminated (or at least not much more so than other ultracool dwarf populations) by binaries whose combined spectra mimic those of a bonafide L9-T4 dwarfs.  On the other hand, we found the unresolved L9-T4 binary fraction to be double that of a primary-spectral-type-selected sample ($6^{+6}_{-4}$\%), which may hint that binaries indeed make up a larger fraction of L9-T4 unresolved spectral types.
In this case, the actual unresolved L9-T4 binary frequency could be on the high end of our inferred distribution for $\nu$ (e.g. $\sim$20\% at the 1$\sigma$ upper limit). A larger sample will be required to resolve these contradictory indications.

\citet{burgasser10} asked the question of whether L/T transition binaries may be identified from their NIR spectra.  This would be advantageous as it would allow us to improve our statistical studies and to identify binary contaminants in the L/T transition regime.    The authors compared the NIR spectra of L/T transition dwarfs to a series of single and composite spectral templates and identified objects with spectral features common to known binaries as ``strong'' and ``weak'' binary candidates depending on the number of common traits.  Our sample contained 2 ``strong'' binary candidates (2M0247+16 and 2M0351+48) and 2 ``weak'' binary candidates (2M0119+24 and 2M0758+32) suggested by \citet{burgasser10}, and none were resolved into multiples.  This result suggests that either (i)we cannot identify binaries reliably using spectral indices, or (ii)that the estimated number of binaries missed at low separations is significant.  In all likelihood both of these explanations are partially true.  In the first case, spectral irregularities in the candidate binaries must be caused by atypical atmospheric or physical properties, rather than binarity.  This is likely the case for another strong candidate binary of \citet{burgasser10}, 2M2139+02, which was found to have peculiar atmospheric characteristics \citep[patchy clouds, and large-amplitude photometric variability;][]{radigan12}.  Alternatively, the second explanation may find support in radial velocity surveys of VLM stars and BDs both in the field or in young clusters  \citep[e.g.][]{maxted05,basri06, joergens08} which show that there may be just as many binaries at at separations $<$ 3\,AU as found by direct imaging surveys.  In this latter case the true L9-T4 binary frequency could be as high as $\sim$30-40\%.  Even so, this would imply that the majority ($\sim$2/3) of objects in the late-L/T transition are single.
 
\acknowledgements
The authors wish to thank the Keck support astronomers, and in particular Randy Campbell and Hien Tran, for assistance in conducting the NIRC2 and OSIRIS observations presented here.  JR is supported in part by a
Vanier Canada Graduate Scholarship from the National Sciences and Engineering Research Council of Canada.
RJ is supported in part by research grants from the Natural Sciences and Engineering
Research Council of Canada.  This research has benefited from the SpeX Prism
Spectral Libraries, maintained by Adam Burgasser at
http://pono.ucsd.edu/$\sim$adam/browndwarfs/spexprism.
This publication makes use of data products from the Two Micron All Sky Survey, which is a joint project of the University of Massachusetts and the Infrared Processing and Analysis Center/California Institute of Technology, funded by the National Aeronautics and Space Administration and the National Science Foundation.  Funding for the SDSS and SDSS-II has been provided by the Alfred P. Sloan Foundation, the Participating Institutions, the National Science Foundation, the U.S. Department of Energy, the National Aeronautics and Space Administration, the Japanese Monbukagakusho, the Max Planck Society, and the Higher Education Funding Council for England. The SDSS Web Site is http://www.sdss.org/.
The authors wish to recognize and acknowledge the very significant cultural role and reverence that the summit of Mauna Kea has always had within the indigenous Hawaiian community.  We are most fortunate to have the opportunity to conduct observations from this mountain.

\appendix
\section{A. The Empirical Flux Ratio Distribution of Resolved L and T Dwarf Binaries}
\begin{figure*}
\epsscale{0.7}
\plotone{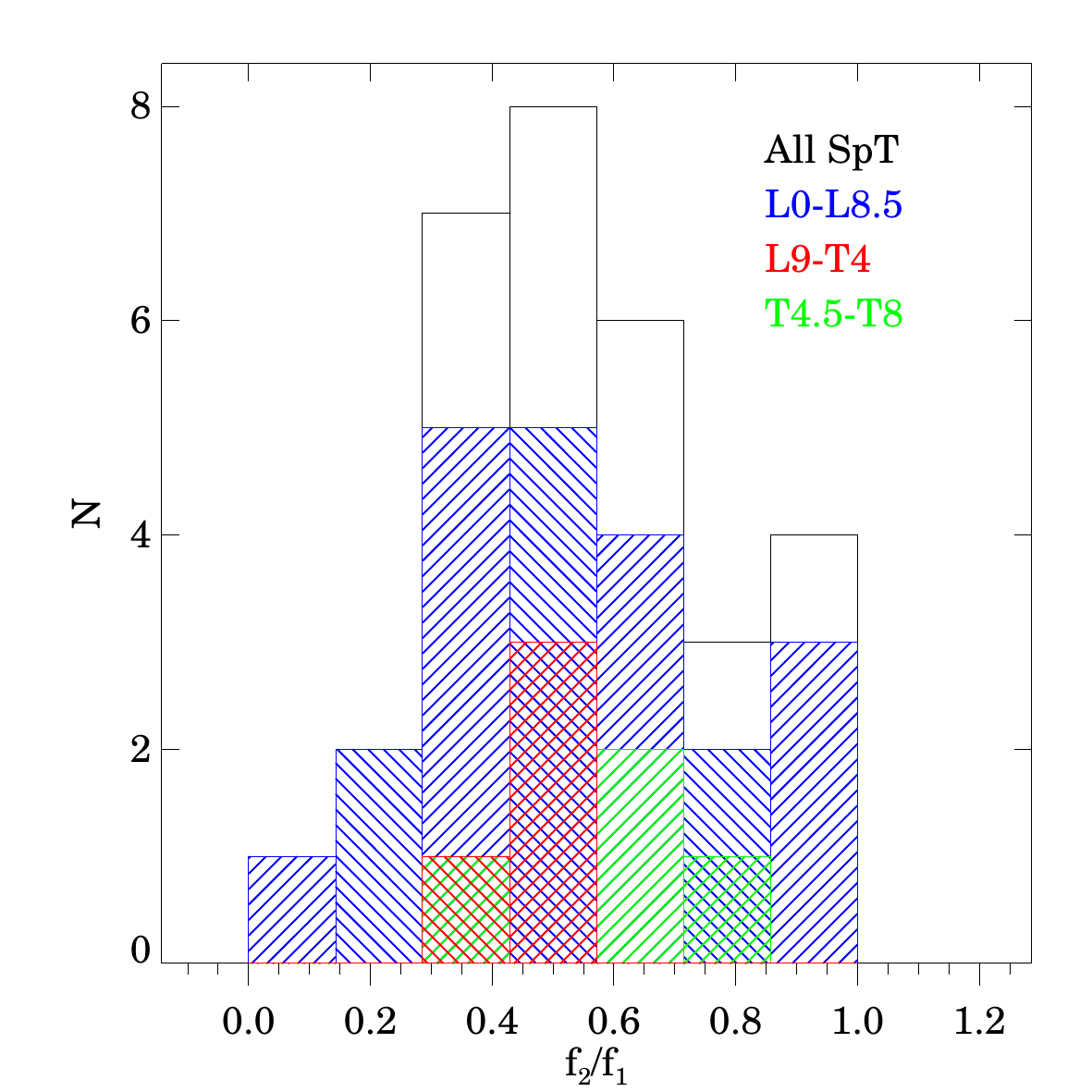} 
\caption{The distribution of flux ratios for resolved L and T dwarf binaries in table \ref{tab:fr} (black histogram). Colored bars show how the sample subdivides by unresolved spectral type.  Integrating over this empirical distribution using equation \ref{eq:alpha}, we find $\alpha$=1.87, which is equivalent to the value for for a flat distribution, and strongly differs from the value found by assuming flux ratios peak at 1. \label{fig:fr}}
\end{figure*}

In order to convert the observed binary fraction into a volume-corrected binary frequency using equation \ref{eq:bin} we needed to compute the ratio of volume searched for binaries over single objects, $\alpha=(d_{\rm bin}/d_{\rm single})^3$.  To do this we used equation \ref{eq:alpha} which is originally given in \citet{burgasser03}, and depends on the distribution of flux ratios for resolved binaries, ($f_2/f_1$).  The contrasts, $\Delta m$ for known binaries reported by \citet{burgasser03,bouy03,burgasser06,reid06,reid08} are given in table \ref{tab:fr}. We converted these contrasts to flux ratios according to $f_2/f_1=10^{-0.4*\Delta m}$.  This empirical flux ratio distribution is shown in figure \ref{fig:fr}.  Integrating directly over this empirical distribution in equation \ref{eq:bin} we found $\alpha=1.87$, which is almost identical to the value obtained for a flat flux ratio distribution of $\alpha=1.86$.

\bibliographystyle{/home/radigan/manuscripts/astronat/apj/apj}
\bibliography{/home/radigan/manuscripts/lib2}

\clearpage

\clearpage

\clearpage

 \begin{figure*}
\epsscale{0.7}
\plotone{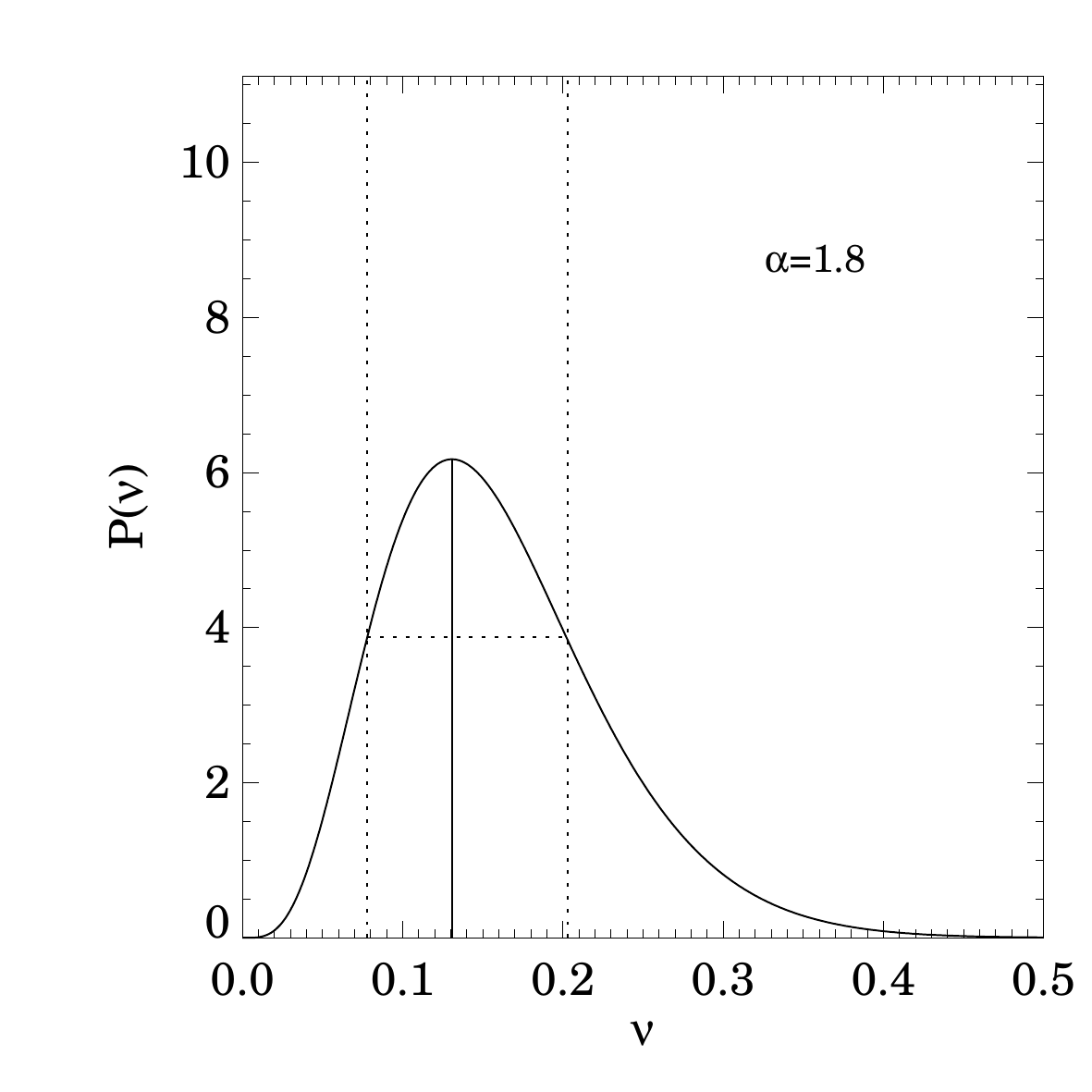}
\caption{Probability distribution for the volume-bias corrected L9-T4 binary frequency at projected separations $\gtrsim 1-2.5$\, AU.  Dotted lines show the 68\% credible region about the most probable value. \label{fig:binom}}
\end{figure*}

\clearpage

\begin{deluxetable*}{lcccccc}
\tablecaption{Targets \label{tab:targets}}
\tablecolumns{7}
\tablewidth{6in}
\tablehead{
\colhead{Target ID} & 
\colhead{SpT} & 
\colhead{J} &
\colhead{H} &
\colhead{K$_s$} &
\colhead{J-K$_s$} &
\colhead{Reference(s)\tablenotemark{a}}}

\startdata
2MASS J01191207+2403317 & T2 & 17.0 & 16.0  & 17.0 & 1.0  & 1 \\
2MASS J02474978-1631132 & T2 & 17.2 & 16.2  & 15.6 & 1.57 & 1 \\
2MASS J03284265+2302051 & L8 &16.7  & 15.6 & 14.9 & 1.8  & 2, 4 \\
2MASS J03510423+4810477 & T1 & 16.5 & 15.6 & 15.0 & 1.50  & 1 \\
2MASS J06020638+4043588 & T4.5 & 15.5 & 15.6 & 15.2 & 0.37  & 3\\
2MASS J07420130+2055198 & T5 & 16.2 &  15.9 & ~~16.2\tablenotemark{a} & 0.00  & 4, 5\\
2MASS J07584037+3247245 & T2 & 15.0 & 14.1 & 13.9 & 1.07  & 4, 5\\
2MASS J08381155+1511155 & T3 & 16.65 & 16.21 & 16.2\tablenotemark{b}  & 0.45 &  6, 7
\enddata
\tablenotetext{a}{Discovery reference as well as the NIR spectral type reference if different form the former.}
\tablenotetext{b}{undetected in 2MASS, derived from spectrum}
\tablerefs{(1)\citet{chiu06} (2)\citet{kirkpatrick00} (3)\citet{looper07} (4)\citet{knapp04} (5)\citet{burgasser06} (6)\citet{aberasturi11} (7)This paper}
\end{deluxetable*} 

\begin{deluxetable*}{lccccccccc}
\tablecaption{Observations\label{tab:obs}}
\tablecolumns{10}
\tablewidth{7in}
\tablehead{
\colhead{Target} & 
\colhead{SpT} & 
\colhead{Filter} & 
\colhead{N$_{exp}$} &
\colhead{t$_{exp}$ (s)} &
\colhead{airmass\tablenotemark{a}} &
\colhead{FWHM (\arcsec)} &
\colhead{Strehl} &
\colhead{TT R Mag \tablenotemark{b}} &
\colhead{TT Sep \tablenotemark{c}}
}
\startdata
2M0119+24 & T2 &  $K_s$ & 3 & 120 & 1.03 & 0.065 &  0.23 & 16.0 & 9.0 \\
2M0247$-$16 & T2 & $K_s$ & 5 & 80 &  1.24 & 0.073 & 0.19 & 13.8 & 45.7 \\
2M0328+23 & L8 & $K_s$ & 6 & 100 &  1.01 & 0.085 & 0.15 & 16.9 & 61.2 \\
2M0351+48 & T1 &  $K_s$ & 5 & 80 & 1.15 & 0.077 & 0.17 & 10.0 & 55.0 \\
2M0602+40 & T4.5 & $K_s$ & 3 & 60 & 1.1 & 0.056 & 0.31 & 10.6 & 50.9 \\
2M0742+20 & T5 & $K_s$ & 2 & 80 & 1.16 & 0.087 & 0.17 &  16.0 & 38.2 \\
2M0758+32 & T2 &  $K_s$ & 6 & 30 & 1.13 & 0.075 & 0.20 & 15.9 & 40.9  \\
2M0838$+$15 & T3.5 & $K_s$ & 4 & 120 & 1.18 & 0.081 & 0.25 & 12.6 & 51.1  \\
2M0838$+$15 & T3.5 & $J$ & 4 & 210 & 1.13 & 0.106 & 0.05 & 12.6 & 51.1  \\
2M0838$+$15 & T3.5 & $H$ & 4 & 180 & 1.10 & 0.083 & 0.13 & 12.6 & 51.1  
\enddata
\tablenotetext{a}{Air mass at the start of the observation}
\tablenotetext{b}{R magnitude of the tip-tilt star}
\tablenotetext{c}{Angular distance between the target and tip-tilt star}
\end{deluxetable*} 

\clearpage

\begin{deluxetable}{cccccccccccc}
\label{tab:mpfit}
\tabletypesize{\scriptsize}
\tablecaption{\redcol $BC$ System Fit Parameters}
\tablecolumns{12}
\tablewidth{6.5in}
\tablehead{
\multicolumn{9}{c}{{\bf Best fit parameters for BC components}} & \multicolumn{3}{c}{{\bf Fit information for  component A}} \\
\colhead{Filter} &  \colhead{No.} & \colhead{FWHM} & \colhead{Strehl} &\colhead{Sep (mas)} & \colhead{PA (deg)} & \colhead{$F_B/F_A$} & \colhead{$F_C/F_A$} & \colhead{$\chi^2$/dof} & \colhead{$N_{G}$} & \colhead{$\chi^2$/dof}  & \colhead{BIC}
}
\startdata
$J$& 1 & 112 &0.032 & 48.1$\pm$0.8 &-14$\pm$1 & 0.60$\pm$0.03 &0.75$\pm$0.03 & 1.68  &  5 &  1.84 & 2987  \\
$J$& 2 & 81 &0.061 & 49.8$\pm$0.5 &-2$\pm$1 & 063$\pm$0.01 &0.73$\pm$0.01 & 1.25 & 5 & 1.62 &  2652 \\
$J$& 3 & 109 &0.041 & 48.5$\pm$0.7 &-9$\pm$1 &0.61$\pm$0.02 &0.76$\pm$0.02  & 1.24 & 7 & 1.59 & 2517 \\
{\bf Mean} &  &  & &  {\bf 49.1} & {\bf -7} & {\bf 0.62} & {\bf 0.74} & & & & \\  
{\bf SD} &  &  &  &  {\bf 0.9} & {\bf 6} & {\bf 0.02} & {\bf 0.01} &  &  & & \\  
\hline
$H$& 1 & 98 & 0.096 & 49.7$\pm$0.4  &-9$\pm$1 & 0.77$\pm$0.01&0.62$\pm$0.01 & 1.15 & 5 & 1.24 & 2081 \\
$H$& 2 & 86 & 0.129 & 50.8$\pm$0.3 &- 4$\pm$1& 0.75$\pm$0.01 & 0.63$\pm$0.01 & 1.11 & 6 & 0.94 & 1677 \\
$H$& 3 & 80 & 0.123 & 50.2$\pm$0.2 &-5$\pm$1 & 0.73$\pm$0.01&0.66$\pm$0.01 & 1.15 & 5 & 1.09 & 1861 \\
$H$& 4 & 67 & 0.172 & 50.1$\pm$0.2 &-6$\pm$1 & 0.74$\pm$0.01&0.64$\pm$0.01 & 1.11 & 7 & 1.20 & 2100 \\
$H$& 5 & 73 & 0.146 & 50.0$\pm$0.3 &-6$\pm$1 & 0.71$\pm$0.01 & 0.65$\pm$0.01 & 1.21 & 7 & 1.20 & 2078 \\
$H$& 6 & 95 & 0.095 & 49.7$\pm$0.4 &-8$\pm$1 &0.73$\pm$0.01&0.69$\pm$0.01 & 0.93 & 4 & 1.16 & 1927 \\
{\bf Mean} &  &  & &  {\bf 50.2} & {\bf -6} & {\bf 0.74} & {\bf 0.65} & & & & \\  
{\bf SD} &  &  &  &  {\bf 0.4} & {\bf 2} & {\bf 0.02} & {\bf 0.02} &  &  & & \\  
\hline
$K_s$& 1 & 89 & 0.213 & 50.1$\pm$0.6 &-6$\pm$2 & 0.71$\pm$0.02&0.63$\pm$0.02 & 1.17 & 3 & 0.99 & 1645 \\
$K_s$& 2 & 85 & 0.245 & 50.8$\pm$0.5 &-1$\pm$1 & 0.71$\pm$0.01&0.59$\pm$0.01 & 0.69 & 3 & 0.71 & 1213 \\
$K_s$& 3 & 74 & 0.302 & 50.8$\pm$0.4 &-7$\pm$1 & 0.75$\pm$0.01 & 0.59$\pm$0.01 & 0.84 & 4 & 0.75 & 1324 \\
$K_s$& 4 & 77 & 0.253 & 51.2$\pm$0.5 &-5$\pm$1 &0.73$\pm$0.01&0.61$\pm$0.01 & 0.83 & 3 & 0.72 & 1230 \\
{\bf Mean} &  &  & &  {\bf 50.9} & {\bf -5} & {\bf 0.73} & {\bf 0.60} & & & & \\  
{\bf SD} &  &  &  &  {\bf 0.5} & {\bf 2} & {\bf 0.02} & {\bf 0.02} &  &  & & 

\enddata
\label{tab:mpfit}
\end{deluxetable} 
\clearpage

\begin{deluxetable}{lcc}
\tablecaption{2M0838$+$15 ABC System Properties}
\tablecolumns{3}
\tablewidth{3in}
\tablehead{
\colhead{Parameter} & 
\colhead{Value} & 
\colhead{Reference}}
\startdata
Identifier &  J08381155+1511155 & 2\\
$\alpha$ (J2000)  & 08$^h$38$^m$11$^s$.55 & 2\\
$\delta$ (J2000)  & +15$^{a}$ 11\arcmin15\arcsec .5  & 2\\
$\mu_{\alpha}\cos{\delta}$ (\arcsec) & $-0.121\pm0.031$ & 3 \\
$\mu_{\delta}$ (\arcsec) & $-0.032\pm0.051$ & 3 \\
\hline
J & 16.65$\pm$0.16& 2 \\
H & 16.21$\pm$0.17 & 2 \\
K$_s$\tablenotemark{a} & 16.20$\pm$0.20 & 1\\
J$-$K$_s$\tablenotemark{a} & 0.45$\pm$0.07 & 1\\
J$-$H\tablenotemark{a} & 0.51$\pm$0.07 & 1\\
W1 & 15.71$\pm$0.07& 3\\
W2 & $14.57$& 3 \\
\hline
NIR SpT\tablenotemark{b} & T3.5  & 1\\
d (pc)\tablenotemark{c} &  49$\pm12$ & 1 \\
${\rm \rho_{BC}}$ (mas)  & 50.2$\pm$0.5& 1\\
${\rm \theta_{BC}}$ (deg) & -6$\pm$2 & 1\\
${\rm \rho_{A(BC)}}$ (mas) & 548.6 $\pm$1.2 & 1\\
${\rm \theta_{A(BC)}}$ (deg) & 18.8$\pm$0.1 & 1\\
${\rm q_{AB}}$\tablenotemark{d}&  $0.89-0.92$ & 1 \\
${\rm q_{(BC)A}}\tablenotemark{d}$ & 0.56-0.57 & 1 
\enddata
\label{tab:sys_prop}
\tablenotetext{a}{Synthetic 2MASS photometry from SpeX spectrum}
\tablenotetext{b}{Unresolved spectral type}
\tablenotetext{c}{Spectroscopic parallax}
\tablenotetext{d}{Range or value provided spans values inferred for 0.3-3\,Gyr}
\tablerefs{(1) This work; (2)2MASS Point Source Catalog \citep{2mass}; (3)\citet{aberasturi11}}
\end{deluxetable}

\begin{deluxetable}{lccc}
\tablecaption{2M0838$+$15 ABC Component Properties}
\tablecolumns{4}
\tablewidth{3 in}
\tablehead{
\colhead{Parameter} & \colhead{$A$} & \colhead{$B$} & \colhead{C}}
\startdata
NIR SpT & T3$\pm$1 & T3 $\pm$1 & T4.5$\pm$1 \\
J         & 17.57$\pm$0.16 & 18.04$\pm$0.16 & 17.98$\pm$0.16 \\ 
H        & 17.10$\pm$0.17 & 17.43$\pm$0.18 & 17.78$\pm$0.18 \\
K$_s$ & 17.11$\pm$0.28 & 17.46$\pm$0.20 & 17.67$\pm$0.20 \\
J$_{\rm MKO}$ & 17.34$\pm$0.16 & 17.81$\pm$0.16 & 17.72$\pm$0.16 \\
$\Delta J_{MKO}$  & \nodata & 0.52$\pm$0.04 & 0.38$\pm$0.04 \\
$\Delta$H & \nodata & 0.33$\pm$0.03 & 0.47$\pm$0.05 \\
$\Delta$K$_s$\tablenotemark{a} & \nodata & 0.35$\pm$0.02 & 0.56 $\pm$0.04 \\
$M$ (3 Gyr, $M_{\odot}$) & 0.060$\pm$0.009 & 0.055$\pm$0.008 & 0.050$\pm$0.008 \\
$M$ (300 Myr, $M_{\odot}$) \tablenotemark{b}  & 0.020$\pm$0.005 & 0.018$\pm$0.005 & 0.016$\pm$0.004
\enddata
\label{tab:comp_prop}
\tablenotetext{a}{Differential magnitude with respect to component A}
\tablenotetext{b}{\redcol Masses at 300 Myr are multi-valued (see figure \ref{fig:iso}).  The high-mass solution for the 300 Myr isochrone is provided here. }
\end{deluxetable} 

\begin{deluxetable*}{lccccccccc}
\tablecaption{The Composite L9-T4 Sample}
\tablecolumns{10}
\tablewidth{6in}
\tablehead{
\colhead{2MASS ID} & 
\colhead{SpT\tablenotemark{a}} & 
\colhead{$J$} &
\colhead{$J-K_s$} &
\colhead{$m-M$} &
\colhead{$d$ (pc)} &
\colhead{$\theta$(\arcsec)} &
\colhead{Ref\tablenotemark{b}} &
\colhead{Ref \tablenotemark{c}}
}
\startdata
2MASS J01191207$+$2403317 & T2 & $17.02\pm0.18$ & $1.00\pm0.26$ & $2.51\pm0.50$ & $~31.8\pm6.6$\tablenotemark{c} & \nodata &  4  & \nodata \\
2MASS J01365662$+$0933473 & T2.5 & $13.45\pm0.03$ & $0.89\pm0.04$ & $-1.07\pm0.46$~ & $~~6.1\pm1.3$\tablenotemark{c} & \nodata & 3  & \nodata \\
2MASS J01514155$+$1244300 & T1 & $16.57\pm0.13$ & $1.38\pm0.23$ & $1.92\pm0.50$ & $21.4\pm1.5$ & \nodata & 2  & 7 \\
2MASS J02474978$-$1631132 & T2 & $17.19\pm0.18$ & $1.57\pm0.27$ & $2.11\pm0.50$ & $~26.5\pm5.5$\tablenotemark{c} & \nodata & 4  & \nodata \\
2MASS J03284265$+$2302051 & L9.5 & $16.69\pm0.14$ & $1.78\pm0.18$ & $1.97\pm0.47$ & $30.2\pm3.8$ & \nodata & 10,1,5 & 7 \\
2MASS J03510423$+$4810477 & T1 & $16.47\pm0.13$ & $1.47\pm0.18$ & $1.73\pm0.48$ & $~22.2\pm4.6$ \tablenotemark{c}& \nodata  & 4  & \nodata \\
2MASS J04234858$-$0414035 & T0 & $14.47\pm0.03$ & $1.54\pm0.04$ & $-0.12\pm0.46~$ & $13.9\pm0.2$ & 0.164 & 2  & 8 \\
2MASS J05185995$-$2828372 & T1 & $15.98\pm0.10$ & $1.82\pm0.12$ & $0.90\pm0.47$ & $22.9\pm0.4$ & 0.051 & 2  & 8 \\
2MASS J07584037$+$3247245 & T2 & $14.95\pm0.04$ & $1.07\pm0.07$ & $0.38\pm0.46$ & $~11.9\pm2.5$ \tablenotemark{c}& \nodata  & 4  & \nodata \\
2MASS J08371718$-$0000179 & T1 & $17.10\pm0.21$ & $1.23\pm0.30$ & $2.61\pm0.51$ & $~30\pm12$ & \nodata  & 2  & 7 \\
2MASS J09083803$+$5032088 & L9 & $14.55\pm0.02$ & $1.60\pm0.04$ & $0.10\pm0.46$ & $~10.5\pm2.2$ \tablenotemark{c}& \nodata & 5  & \nodata \\
2MASS J09201223$+$3517429 & T0p & $15.62\pm0.06$ & $1.65\pm0.09$ & $0.93\pm0.46$ & $29.1\pm0.7$ & 0.075 & 11,1  & 8 \\
2MASS J10210969$-$0304197 & T3 & $16.25\pm0.09$ & $1.13\pm0.20$ & $1.35\pm0.49$ & $33.4\pm1.5$ & 0.172 &  2  & 8 \\
2MASS J12545393$-$0122474 & T2 & $14.89\pm0.04$ & $1.05\pm0.06$ & $0.33\pm0.46$ & $11.8\pm0.3$ & \nodata & 2  & 6 \\
2MASS J14044941$-$3159329 & T2.5 & $15.58\pm0.06$ & $1.04\pm0.11$ & $0.90\pm0.47$ & $23.8\pm0.6$ & \nodata & 3  & 8 \\
2MASS J17503293$+$1759042 & T3.5 & $16.34\pm0.10$ & $0.86\pm0.21$ & $1.54\pm0.50$ & $27.6\pm3.5$ & \nodata & 2  & 7 \\
2MASS J20474959$-$0718176 & T0 & $16.95\pm0.20$ & $1.57\pm0.28$ & $2.33\pm0.50$ & $20.0\pm3.2$ & \nodata & 3  & 9 \\
2MASS J22541892$+$3123498 & T4 & $15.26\pm0.05$ & $0.36\pm0.15$ & $0.80\pm0.48$ & $~14.4\pm3.0\tablenotemark{c}$ & \nodata & 2  & \nodata
\enddata
\label{tab:sample}
\tablerefs{(1)\citet{bouy03} (2)\citet{burgasser06} (3)\citet{goldman08} (4)This paper (5)\citet{reid06} (6)\citet{dahn02}  (7)\citet{vrba04}  (8)\citet{dupuy12}  (9)\citet{faherty12} (10)\citet{gizis03} (11)\citet{reid01}}
\tablenotetext{a}{Unresolved NIR Spectral Type}
\tablenotetext{b}{Survey References}
\tablenotetext{c}{Parallax References.  Distances without parallaxes are based on SpT vs absolute $K_s$ magnitude relationship of \citet{dupuy12}.}
\end{deluxetable*}

\begin{deluxetable*}{lccccccccc}
\tablecaption{Flux ratios of resolved binaries}
\tablecolumns{10}
\tablewidth{6in}
\tablehead{
\colhead{2MASS ID} & 
\colhead{SpT\tablenotemark{a}} & 
\colhead{SpT A} &
\colhead{SpT B} &  
\colhead{$\Delta m$} &
\colhead{$f_2/f_1$} &
\colhead{Sep (mas)} &
\colhead{Filter} &
\colhead{Ref. 1\tablenotemark{b}} &
\colhead{Ref. 2\tablenotemark{c}}
}
\startdata
2MASS J00043484$-$4044058 & L5 & L5 & L5 & 0.10 & 0.91 & 90 & F110W & 1 &  5 \\
2MASS J00250365$+$4759191 & L4 & L4 & L4 & 0.17 & 0.86 & 330 & F110W & 1 &  1 \\
2MASS J02052940$-$1159296 & L7 & L7 & L7 & 0.63 & 0.56 & 190 & F814W & 2 &  6 \\
2MASS J03572695$-$4417305 & L0 & M9 & L1.5 & 1.50 & 0.25 & 97 & F814W & 2 &  7 \\
2MASS J04234858$-$0414035 & T0 & L6.5 & T2 & 0.82 & 0.47 & 164 & F170M & 3 &  5 \\
2MASS J05185995$-$2828372 & T1 & L6 & T4 & 0.90 & 0.44 & 51 & F170M & 3 &  5 \\
2MASS J07003664$+$3157266 & L3.5 & L3 & L6.5 & 1.20 & 0.33 & 170 & F110W & 1 &  5 \\
2MASS J07464256$+$2000321 & L0.5 & L0 & L1.5 & 1.00 & 0.40 & 146 & F814W & 8,2 &  5 \\
2MASS J08503593$+$1057156 & L6 & L6.5 & L8.5 & 1.47 & 0.26 & 264 & F814W & 8,2 &  5 \\
2MASS J08564793$+$2235182 & L3 & L3 & L9.5 & 2.76 & 0.08 & 374 & F814W & 9,2 &  6 \\
2MASS J09153413$+$0422045 & L7 & L7 & L7.5 & 0.12 & 0.90 & 730 & F110W & 1 &  6 \\
2MASS J09201223$+$3517429 & T0p & L5.5 & L9 & 0.88 & 0.44 & 219 & F814W & 8,2 &  5 \\
2MASS J09261537$+$5847212 & T4.5 & T3.5 & T5 & 0.40 & 0.69 & 70 & F170M & 3 &  5 \\
2MASS J10170754$+$1308398 & L2 & L1.5 & L3 & 0.74 & 0.51 & 75 & F1042 & 9,2 &  5 \\
2MASS J10210969$-$0304197 & T3 & T0 & T5 & 1.03 & 0.39 & 172 & F170M & 3 &  5 \\
2MASS J11122567$+$3548131 & L4.5 & L4.5 & L6 & 1.04 & 0.38 & 294 & F1042 & 9,2 &  5 \\
2MASS J11463449$+$2230527 & L3 & L3 & L3 & 0.75 & 0.50 & 102 & F814W & 8,2 &  5 \\
2MASS J12255432$-$2739466 & T6 & T6 & T8 & 1.05 & 0.38 & 282 & F1042 & 4 &  3 \\
2MASS J12281523$-$1547342 & L5 & L5.5 & L5.5 & 0.40 & 0.69 & 70 & F814W & 2 &  5 \\
2MASS J12392727$+$5515371 & L5 & L5 & L6 & 0.54 & 0.61 & 252 & F1042 & 9,2 &  6 \\
2MASS J14304358$+$2915405 & L2 & L2 & L3.5 & 0.45 & 0.66 & 157 & F1042 & 9,2 &  6 \\
2MASS J14413716$-$0945590 & L0.5 & L0.5 & L1 & 0.34 & 0.73 & 83 & F814W & 2 &  6 \\
2MASS J14493784$+$2355378 & L0 & L0 & L3 & 1.08 & 0.37 & 134 & F1042 & 9,2 &  6 \\
2MASS J15344984$-$2952274 & T5.5 & T5.5 & T5.5 & 0.20 & 0.83 & 110 & F1042 & 4 &  3 \\
2MASS J15530228$+$1532369 & T7 & T6.5 & T7.5 & 0.46 & 0.65 & 349 & F170M & 3 &  5 \\
2MASS J16000548$+$1708328 & L1.5 & L1.5 & L1.5 & 0.69 & 0.53 & 57 & F814W & 9,2 &  6 \\
2MASS J17281150$+$3948593 & L7 & L5 & L7 & 0.66 & 0.54 & 131 & F814W & 9,2 &  5 \\
2MASS J21011544$+$1756586 & L7.5 & L7 & L8 & 0.59 & 0.58 & 234 & F814W & 9,2 &  5 \\
2MASS J21522609$+$0937575 & L6 & L6 & L6 & 0.15 & 0.87 & 250 & F110W & 1 &  6 \\
2MASS J22521073$-$1730134 & L7.5 & L4.5 & T3.5 & 1.12 & 0.36 & 140 & F110W & 1 &  5
\enddata
\label{tab:fr}
\tablenotetext{a}{Unresolved NIR Spectral Type}
\tablenotetext{b}{Survey References}
\tablenotetext{c}{Resolved spectral type references.}
\tablerefs{(1)\citet{reid06} (2)\citet{bouy03} (3)\citet{burgasser06} (4)\citet{burgasser03} (5)\citet{dupuy12}  (6)Trent J. Dupuy, Private Comm. (7)\citet{martin06} (8)\citet{reid01} (9)\citet{gizis03}}
\end{deluxetable*}

\end{document}